\documentclass[prd,nofootinbib,preprint,superscriptaddress]{revtex4}
\pdfoutput=1

\usepackage{amsmath,amssymb}
\usepackage{epsfig}
\usepackage{graphicx}
\usepackage[usenames,dvipsnames]{color}
\usepackage{subfigure}
\usepackage{slashed}
\usepackage[colorlinks,citecolor=blue]{hyperref}
\usepackage{color}
{\newcommand{\lsim}{\mbox{\raisebox{-.6ex}{~$\stackrel{<}{\sim}$~}}}
	{\newcommand{\gsim}{\mbox{\raisebox{-.6ex}{~$\stackrel{>}{\sim}$~}}}

\begin{document}
\title{Connecting low scale seesaw for neutrino mass to inelastic sub-GeV dark matter with Abelian gauge symmetry}
	
	\author{Debasish Borah}
	\email{dborah@iitg.ac.in}
	\affiliation{Department of Physics, Indian Institute of Technology Guwahati, Assam 781039, India}
	
	\author{Satyabrata Mahapatra}
	\email{ph18resch11001@iith.ac.in}
	\affiliation{Department of Physics, Indian Institute of Technology Hyderabad, Kandi, Sangareddy 502285, Telangana, India}
	
	\author{Narendra Sahu}
	\email{nsahu@phy.iith.ac.in}
	\affiliation{Department of Physics, Indian Institute of Technology Hyderabad, Kandi, Sangareddy 502285, Telangana, India}
	
\begin{abstract}
Motivated by the recently reported excess of electron recoil events by the XENON1T experiment, we propose low scale seesaw scenarios for light neutrino masses within $U(1)_X$ gauge extension of the standard model 
that also predicts stable as well as long lived dark sector particles. The new fields necessary for seesaw realisation 
as well as dark matter are charged under the $U(1)_X$ gauge symmetry in an anomaly free way. A singlet scalar field which 
effectively gives rise to lepton number violation and hence Majorana light neutrino masses either at tree or radiative 
level, also splits the dark matter field into two quasi-degenerate states. While sub-eV neutrino mass and non-zero dark 
matter mass splitting are related in this way, the phenomenology of sub-GeV scale inelastic dark matter can be very rich 
if the mass splitting is of keV scale. We show that for suitable parameter space, both the components with keV splitting 
can contribute to total dark matter density of the present universe, while opening up the possibility of the heavier dark 
matter candidate to undergo down-scattering with electrons. We check the parameter space of the model for both fermion and 
scalar inelastic dark matter candidates which can give rise to the XENON1T 
excess while being consistent with other phenomenological bounds. We also discuss the general scenario where mass 
splitting~$\Delta m$ between the two dark matter components can be larger, effectively giving rise to a single component dark 
matter scenario.
\end{abstract}	
\maketitle
	
\section{Introduction}\label{intro}
Recently, the XENON1T collaboration has reported an excess of electron recoil events over the background in the recoil energy $E_r$ in a range 1-7 keV, 
peaked around 2.4 keV\cite{Aprile:2020tmw}. Though this excess is consistent with the solar axion model at $3.5\sigma$ significance and 
with neutrino magnetic moment signal at $3.2\sigma$ significance, both these interpretations face stringent stellar cooling bounds. Since 
XENON1T collaboration has neither confirmed nor ruled out the possible origin of this excess arising due to beta decay from a small amount of tritium 
present in the detector, it has created a great interest in the particle physics community to search for possible new physics interpretations 
for this excess of electron recoil events. New interpretations for this XENON1T anomaly have been proposed by several authors; for example 
see \cite{Smirnov:2020zwf,Takahashi:2020bpq,Alonso-Alvarez:2020cdv,Kannike:2020agf,Fornal:2020npv,Du:2020ybt,Su:2020zny,Harigaya:2020ckz, Borah:2020jzi, Bramante:2020zos, Bell:2020bes, Bally:2020yid}. Among the WIMP type candidates as origin of this excess, the light boosted DM or light inelastic DM are the promising ones.

In this work, we try to connect low scale seesaw origin of non-zero masses of light neutrinos with sub-GeV inelastic DM within an abelian gauge 
extension of the SM. Recently, several low scale seesaw models with additional $U(1)_X$ gauge symmetries have been proposed in different contexts; for 
example, see \cite{Bertuzzo:2018ftf, Camargo:2018uzw, Ballett:2019cqp, Gehrlein:2019iwl, Abdullahi:2020nyr} and references therein. Here we extend the 
SM with a gauged $U(1)_X$ symmetry which primarily describes the dark sector of our model comprising of inelastic DM candidates and fields responsible 
for seesaw realisation. The breaking of $U(1)_X$ to a remnant $Z_2$ symmetry gives rise sub-eV masses of light neutrinos by generating a dimension five 
effective operator $\mathcal{O}_1 LLH H/\Lambda$~\cite{Weinberg:1979sa} at low energy, where $L$ and $H$ are lepton and Higgs doublets respectively, 
$\mathcal{O}_1$ is the Wilson coefficient that depends upon different couplings of the UV complete theory and $\Lambda$ is the scale of $U(1)_X$ symmetry 
breaking. Note that this operator breaks lepton number by two units and hence the corresponding neutrino mass is of Majorana type. After the electroweak 
phase transition the neutrino mass is given by: $m_\nu \propto \langle H \rangle^2/\Lambda$. The breaking of $U(1)_X$ gauge symmetry to a discrete 
$Z_2$ also lead to a small mass splitting between the components of either a Dirac fermion DM or a complex scalar DM, leading to an inelastic DM 
scenario. We ensure the stability of DM via the remnant $Z_2$ symmetry. If we assume that the DM mass is of sub-GeV scale with mass splitting of 
keV or smaller, then both the components can be present today and can give rise interesting phenomenology. In particular, if the mass splitting is 
a few keV, then we can have a tantalising scenario where the heavier component scatters off the electron and gets converted to the lighter component 
effectively explaining the XENON1T anomaly~\cite{Aprile:2020tmw}. We first discuss viable low scale seesaw models within an $U(1)_X$ gauge symmetric 
framework which also predicts inelastic DM. Since all such seesaw models have similar DM phenomenology, we study the latter for a sub-GeV inelastic 
DM whose interactions with the SM relies primarily on kinetic mixing of $U(1)_X$ gauge symmetry with $U(1)_Y$ of the SM. We calculate the relic 
abundance of DM and constrain the parameter space from all available bounds and the requirement of fitting XENON1T excess. While inelastic DM and its connection to origin of light neutrino masses have been studied in earlier works too (see \cite{Gu:2018kmv} for example and references therein), most of these works focused on heavy DM regime with masses in the range of electroweak scale to a few TeV. Such DM scenarios are probed typically at nuclear recoil experiments. In this work, we focus on sub-GeV mass regime of inelastic DM in the context of electron recoil experiments while showing possible connections to the origin of light neutrino mass in different neutrino mass models.

The paper is organised as follows. In section \ref{sec2}, we discuss different low scale seesaw models for light neutrino mass within a 
framework $U(1)_X$ gauge symmetry which also gives rise to inelastic DM. In section \ref{sec:constraint} we discuss the constraints on such low 
scale $U(1)_X$ gauge models followed by detailed discussion of inelastic DM with sub-GeV mass and its implications for XENON1T excess in 
section \ref{inelastic_dm}. We finally conclude in section \ref{sec4}.

\section{Low scale dark seesaw with $U(1)_X$ gauge symmetry }\label{sec2}
In this section, we describe different seesaw realisations for sub-eV masses of light neutrinos by considering the presence of an Abelian gauge 
symmetry $U(1)_X$ which plays a crucial role in both neutrino and dark matter sectors. In particular, the dark matter phenomenology and light neutrino masses are correlated in the following ways.
\begin{itemize}
\item Degenerate two component dark matter implies vanishing light neutrino masses and hence disallowed from experimental data.
\item Tiny mass splitting of keV order or below leads to a two component dark matter scenario with the heavier dark matter candidate being long lived on cosmological scales. The mass splitting of keV scale is of particular interest from XENON1T excess point of view.
\item Larger mass splitting between the DM candidates make heavier DM unstable leading to a single component DM. This, although can not explain XENON1T excess, remains consistent with other DM and neutrino requirements. 
\end{itemize}
We will discuss the second and third scenarios mentioned above one by one. We adopt a minimalistic approach and consider only the newly 
introduced fermions and scalars to be charged under the $U(1)_X$ gauge symmetry leaving the SM particles to be charge-less under this new symmetry. 
Additional discrete symmetries are also incorporated to obtain the desired couplings in the Lagrangian for seesaw realisations. With non-minimal 
particle content one can also consider scenarios with similar seesaw realisation and DM phenomenology without any additional discrete symmetries 
as have been studied in several works, for example, see \cite{Adhikari:2008uc, Borah:2012qr, Adhikari:2015woo, Patra:2016shz, Nanda:2017bmi, Barman:2019aku, Biswas:2019ygr, Nanda:2019nqy, Bhattacharya:2020wra, Mahapatra:2020dgk} and references therein.
	
\subsection{Inverse seesaw model with inelastic DM}\label{subsec:iss}
Here we consider an inverse seesaw~\cite{Mohapatra:1986bd}, which is a typical low scale model in contrast to the high scale canonical 
seesaw scenarios like type I, type II and type III~\cite{Minkowski:1977sc, GellMann:1980vs, Mohapatra:1979ia, Schechter:1980gr, Mohapatra:1980yp, 
Lazarides:1980nt, Wetterich:1981bx, Schechter:1981cv, Foot:1988aq}. The inverse seesaw is realised in a gauged $U(1)_X$ extension of a two Higgs 
doublet model. The gauge group of the theory is thus given by: $SU(3)_c \otimes SU(2)_L \otimes U(1)_Y \otimes U(1)_X$. An additional discrete $Z_4$ 
symmetry is also imposed to have the correct mass matrix structure of neutral fermions. As shown in table \ref{table1} (particle 
content of inelastic fermion DM with inverse seesaw) and table \ref{table1a} (particle content of inelastic scalar DM with inverse seesaw), the 
new degrees of freedoms apart from a second Higgs doublet $H_2$ are all singlets under the SM gauge group ($SU(3)_c \otimes SU(2)_L \otimes U(1)_Y$). 
The $U(1)_X$ gauge charges of these newly introduced particles are chosen in such a way that give rise to the desired neutrino and DM phenomenology. 
While $N_R, S_R$ are singlet fermions taking part in inverse seesaw, the fields $\Psi_{L, R}$ and $\eta$ are introduced as viable fermion and scalar 
DM candidates respectively. When the singlet scalar $\Phi_2$ acquires a vacuum expectation value (vev), the gauged $U(1)_X$ symmetry breaks down to a remnant 
$Z_2$ symmetry under which the vector-like fermion singlet $\Psi (=\Psi_L +\Psi_R)$ is odd. As a result, $\Psi$ behaves as a 
candidate of fermion DM\footnote{For scalar DM we need another stabilising symmetry as we discuss below}. We show that the spontaneous breaking of $U(1)_X$ gauge symmetry not only generates the 
lepton number violating mass term for inverse seesaw, but also splits the DM (both fermion and scalar) into two quasi-degenerate components. Note that, we consider only single component DM, either fermion or scalar, not both in the same model. We discuss fermion and scalar DM separately to show 
their inelastic nature arising from a scalar field taking part in generating light neutrino masses. It should be noted that the models we 
considered here are anomaly free by the virtue of the assigned gauge charges of the newly introduced fermions.  
	\begin{table}
		\begin{tabular}{|c c|c|c|c|c|c|c|c|}
			\hline
			& & $N_R$ & $S_R$  & $\Phi_1$&$\Phi_2$& $\Psi_{L,R}$ & $H_2$ \\ 
			\hline
			& $SU(2)_{L}$ & 1 & 1 & 1 & 1 & 1 & 2\\
			\hline
			& $U(1)_{X}$ & 1 & -1 & 0 & 2 & -1 & 1\\
			
			\hline
			& $Z_4$ & 1 & i & i & -1 & -i  & 1\\
			\hline
		\end{tabular}   
		\caption{New particles and their quantum numbers under the imposed symmetries for fermion DM realisation.}
		\label{table1}
	\end{table} 
	
\begin{table}
		\begin{tabular}{|c c|c|c|c|c|c|c|c|}
			\hline
			& & $N_R$ & $S_R$  & $\Phi_1$&$\Phi_2$ & $ \eta$ & $H_2$ \\ 
			\hline
			& $SU(2)_{L}$ & 1 & 1 & 1 & 1 & 1 & 2\\
			\hline
			& $U(1)_{X}$ & 1 & -1 & 0 & 2 & -1 & 1\\
			\hline
			& $Z_4$ & 1 & i & i & -1  &-i & 1\\
			\hline
		        & $Z_2$ & 1 & 1 & 1 & 1 & -1 & 1\\
			\hline
			\end{tabular}   
		\caption{New particles and their quantum numbers under the imposed symmetries for scalar DM realisation.}
		\label{table1a}
	\end{table}

The Lagrangian involving the new degrees of freedom consistent with the extended symmetry is given by
	\begin{align}\label{neutrino_lagrangian}
	-\mathcal{L} &\supset  y_\nu \overline{L} \widetilde{H_2} N_R +y_{NS}N_R S_R \Phi_1^{\dagger} +y_S S_R S_R \Phi_2  + {\rm h.c.} + \mathcal{L}_{\rm DM}\,,
	\end{align}
where $\mathcal{L}_{\rm DM}$ describes the Lagrangian for inelastic DM and is discussed below separately for fermion ($\Psi$) and scalar ($\eta$) cases.

The electroweak symmetry is broken when the Higgs doublets $H_1$ and $H_2$ acquire non-zero vevs, while the vevs of $\Phi_1$ and $\Phi_2$ break 
$Z_4 \times U(1)_X$ symmetry of the hidden sector. The scalar fields which acquire non-zero vevs can be represented as 
\begin{align*}
H_{1,2} = \begin{pmatrix}
	h^+_{1,2}  \\
	\frac{(h_{1,2}+v_{1,2}+i h^I_{1,2})}{\sqrt{2}} 
	\end{pmatrix}, \;\; \Phi_{1, 2}=\frac{\phi_{1, 2}+u_{1, 2}+ i\phi^I_{1,2}}{\sqrt{2}}.
\end{align*}
As can be seen from the Lagrangian in equation \ref{neutrino_lagrangian}, the vev of singlet scalar $\Phi_2$ generates the Majorana mass 
term $\mu$ for $S_R$ field which consequently appears as 33-term (entry for third row and third column) of neutral lepton mass matrix given in 
equation \eqref{neutrino_mass_matrix} and hence is responsible for the light neutrino mass generation through inverse seesaw mechanism. Later we shall 
show that the vev of $\Phi_2$ also creates a mass splitting between the DM components (both for fermion and scalar DM models). \\

	\noindent
	{\bf Light Neutrino Masses}:  \\
In the effective theory, the neutral lepton mass matrix can be written in the basis $n=((\nu_L)^c,N_R,S_R)^T$ as
	\begin{equation}
	-\mathcal{L}_{m_{\nu}} = \frac{1}{2} \overline{(n)^c} \mathcal{M_\nu}n+{\rm h.c.}\,,
	\end{equation}
	where $\mathcal{M_\nu}$ has the structure
	\begin{align}\label{neutrino_mass_matrix}
	\mathcal{M_\nu} =\begin{pmatrix}
	0 & m_D & 0 \\
	m^T_D  & 0 & M \\
	0  &   M   & \mu  
	\end{pmatrix} =\begin{pmatrix}
	0 &  \frac{y_\nu v_2}{\sqrt{2}} & 0 \\
	 \frac{y_\nu v_2}{\sqrt{2}}  & 0 &  \frac{y_{_{NS}} u_1}{\sqrt{2}} \\
	0  &    \frac{y_{_{NS}} u_1}{\sqrt{2}}   &  \frac{y_{_{S}} u_2}{\sqrt{2}}  
	\end{pmatrix}\,.
	\end{align}
	 Assuming that $\mu << m_D < M$, the light neutrino mass matrix at leading order can be given as:
	\begin{align}
	m_\nu &\simeq m^T_D M^{-1} \mu M^{-1} m_D \nonumber \\ & =(\frac{y^T_\nu v_2}{\sqrt{2}})\frac{1}{M}(\frac{y_S u_2}{\sqrt{2}})\frac{1}{M} (\frac{y_\nu v_2}{\sqrt{2}})
	\end{align}
	 For a typical choice: $m_D \sim 10$ GeV, $M \sim 1$ TeV and $\mu \sim 1$ keV, we get sub eV neutrino mass. \\

\noindent
	{\bf Inelastic fermion dark matter}:  \\
	
We now show how inelastic fermion DM arises in this inverse seesaw model. The particle content for inelastic fermion DM realisation 
is already given in table \ref{table1}. The relevant Lagrangian satisfying $U(1)_X \times Z_4$ symmetry can be written as:
\begin{equation}
\mathcal{L}_{\rm DM} = i \overline{\Psi} \gamma^\mu D_\mu \Psi- M(\overline{\Psi_L} \Psi_R + \overline{\Psi_R} \Psi_L) -(  
y_L \Phi_2 \overline{(\Psi_L)^c} \Psi_L + y_R \Phi_2 \overline{(\Psi_R)^c} \Psi_R +h.c )+\frac{\epsilon}{2} B^{\alpha \beta} Y_{\alpha \beta}  
	\end{equation}
where $D_\mu = \partial_\mu + i g' Z'_\mu$ and $B^{\alpha\beta}, Y_{\alpha \beta}$ are the field strength tensors of 
$U(1)_X, U(1)_Y$ respectively and $\epsilon$ is the kinetic mixing parameter. We note that the kinetic mixing plays a crucial role in giving rise the DM phenomenology.

The scalar singlet $\Phi_2$ acquires a non-zero vev $u_2$ and breaks $U(1)_X$ spontaneously down to a remnant $Z_2$ symmetry under which $\Psi_{L,R}$ are 
odd while all other fields are even. As a result, $\Psi_L$ and $\Psi_R$ combine to give a stable DM candidate in the low energy effective theory. The vev 
of $\Phi_2$ also generates Majorana masses for fermion DM: $m_L=y_L u_2/\sqrt{2}$ and $m_R = y_R u_2/\sqrt{2}$ for $\Psi_L$ and $\Psi_R$ respectively. We 
assume $m_L, m_R << M$. As a result, the Dirac fermion $\Psi=\Psi_L + \Psi_R$ splits into two pseudo-Dirac states $\psi_1$ and $\psi_2$ with masses 
$M_1= M-m_+$ and $M_2=M+m_+$, where $m_\pm=(m_L\pm m_R)/2$. Using the expansion of $\Phi_2$ as defined earlier, the Lagrangian in terms of these physical 
mass eigenstates can be written as

\begin{align}
	\mathcal{L}_{\rm DM} = & \frac{1}{2} \overline{\psi_1} i \gamma^\mu \psi_1 + \frac{1}{2} \overline{\psi_2} i \gamma^\mu \psi_2 - 
\frac{1}{2} M_1 \overline{\psi_1} \psi_1 -\frac{1}{2} M_2 \overline{\psi_2} \psi_2 \nonumber \\& +i g' Z'_\mu \overline{\psi_1} \gamma^\mu \psi_2
+\frac{1}{2} g' Z'_\mu (\frac{m_-}{M})(\overline{\psi_2} \gamma^\mu \gamma^5 \psi_2-\overline{\psi_1} \gamma^\mu \gamma^5 \psi_1) \nonumber \\
& +\frac{1}{2}(y_L \cos^2\theta-y_R\sin^2\theta)\overline{\psi_1}\psi_1\phi_2 +\frac{1}{2}(y_R \cos^2\theta-y_L\sin^2\theta)
\overline{\psi_2}\psi_2\phi_2 + \frac{\epsilon}{2} B^{\alpha \beta} Y_{\alpha \beta}\,, 
\end{align}
where $\sin \theta \approx m_-/M$. The mass splitting between the two mass eigenstates is given by $\Delta m =M_2 - M_1= 2m_+ 
= (y_L+y_R) \frac{u_2}{\sqrt{2}} $.  In order to address the XENON1T anomaly in section IV, we take $\Delta m \sim 2.5$ keV.\\

\noindent
	{\bf Inelastic scalar dark matter}:  \\	
We now turn to show the inelastic nature of scalar DM that arises naturally in this inverse seesaw model. The particle content for 
inelastic scalar DM realisation is already given in table \ref{table1a}. Unlike the case of fermion DM, here $U(1)_X \times Z_4$ symmetry alone is not 
enough to stabilise the scalar DM $\eta$. This is due to the presence of a term $H^{\dagger}_1 H_2 \Phi_1 \eta$ in the Lagrangian, allowed by 
$U(1)_X \times Z_4$ symmetry. Therefore, we impose an additional $Z_2$ symmetry under which $\eta$ is odd while rest of the particles are even 
as mentioned in table \ref{table1a}. The relevant Lagrangian involving $\eta$ and $U(1)_X$ gauge boson can be written as:

\begin{equation}
\mathcal{L}_{\rm DM}=(D_\mu \eta)^\dagger (D^\mu \eta) - m^2_\eta \eta^\dagger \eta - (\mu_{\phi} \Phi_2 \eta \eta + {\rm h.c.}) 
+ \frac{\epsilon}{2} B^{\alpha \beta} Y_{\alpha \beta} \,,
\label{scala_lagn}
\end{equation}
where $D_\mu = \partial_\mu + i g' Z'_\mu$ and $B^{\alpha\beta}, Y_{\alpha \beta}$ are the field strength tensors of $U(1)_X, U(1)_Y$ respectively. 
Here $Z'$ is the $U(1)_X$ gauge boson and $g'$ is the corresponding gauge coupling and $\epsilon$ is the kinetic mixing parameter defined earlier.

The scalar $\Phi_2$ acquires a vev and breaks $U(1)_X$ gauge symmetry spontaneously. We parametrise the scalar singlet DM field $\eta$ as:  
\begin{align*}
\eta=\frac{\eta_1 + i \eta_2}{\sqrt{2}} \,.
\end{align*}
Note that the vev of $\Phi_2$ not only gives mass to $Z'$ gauge boson: $M^2_{Z'}= g'^2 (4 v^2_2)$, but also creates a mass splitting between 
$\eta_1$ and $\eta_2$ which is evident from the following effective Lagrangian obtained by putting the above field parametrisation in 
equation~\eqref{scala_lagn}.
\begin{equation}
-\mathcal{L}_{\rm DM} \supseteq (\frac{1}{2}m^2_{\eta} -\frac{\mu_{\phi} u_2}{\sqrt{2}}) \eta^2_1 + (\frac{1}{2}m^2_{\eta} +\frac{\mu_{\phi} u_2}{\sqrt{2}}) \eta^2_2 \,.
\end{equation}   

Thus the mass splitting between the two states $\eta_1$ and $\eta_2$ is given by $\Delta m^2 = m^2_{\eta_2} -m^2_{\eta_1} = \sqrt{2} \mu_{\phi} u_2$. We assume 
$\Delta m << m_{\eta_{1,2}}$. As a result, the two components of $\eta$, i.e. $\eta_1$ and $\eta_2$ give rise viable inelastic DM candidates.  Because of the kinetic mixing between the $U(1)_X$ gauge boson $Z'$ and the SM $Z$ boson, these DM particles can interact with the SM particles which is evident from the following 
effective Lagrangian:
\begin{equation}
\mathcal{L} \supseteq g'Z'^\mu (\eta_1 \partial_\mu \eta_2 - \eta_2 \partial_\mu \eta_1) + \frac{\epsilon}{2} B^{\alpha \beta} Y_{\alpha \beta} 	
\end{equation}

	\subsection{Type II seesaw with inelastic DM}

Here we consider a variant of type II seesaw along with inelastic DM in a gauged $U(1)_X\times Z_4$ extension of the SM. The relevant fields along with 
their quantum numbers are given in tables \ref{table2} (for inelastic fermion DM realisation) and \ref{table2a} (for inelastic scalar DM realisation). 
Except $\Delta_L$ all other newly introduced fields are singlet under the SM gauge group. The scalar field $\Delta_L$ transforms as a triplet under the $SU(2)_L$ and 
possesses a hypercharge 2. The singlet scalars $\Phi_{2}$ and $\Phi_1$ are used to break the $U(1)_X \times Z_4$ symmetry to a remnant 
$Z_2$ symmetry under which the vector-like fermion $\Psi (=\Psi_L +\Psi_R)$ and the scalar singlet $\eta$ are odd. As a result, $\Psi$ ($\eta$) behaves 
as a candidate of fermion (scalar) DM. Note that, we consider only single component DM, either fermion ($\Psi$) or scalar ($\eta$), not both in the 
same model. Here we separately show the inelastic nature of fermion and scalar DM arising from the breaking of $U(1)_X \times Z_4$ symmetry 
which also lead to a low scale type II seesaw.

\begin{table}
		\begin{tabular}{|c c|c|c|c|c|c|c|c|}
			\hline
			& & $L$ & $e_R$ & $ \Delta_L$   & $\Phi_1$&$\Phi_2$& $\Psi_{L,R}$  \\ 
			\hline
			& $U(1)_{X}$ & 0 & 0 & 0  & 0 & 2 & -1\\
			
			\hline
			& $Z_4$ & $i$ & $i$ & -1 & -1 & 1 & -i  \\
			\hline
		\end{tabular}   
		\caption{ New particles and their quantum numbers under the imposed symmetry for type II seesaw origin of neutrino mass and inelastic fermion DM.}
		\label{table2}
	\end{table} 
	
\begin{table}
		\begin{tabular}{|c c|c|c|c|c|c|c|c|}
			\hline
			& & $L$ & $e_R$ & $ \Delta_L$   & $\Phi_1$&$\Phi_2$&  $ \eta$ \\ 
			\hline
			& $U(1)_{X}$ & 0 & 0 & 0  & 0 & 2 &-1\\
			\hline
			& $Z_4$ & $i$ & $i$ & -1 & -1 & 1 & -i \\
			\hline
		\end{tabular}   
		\caption{ New particles and their quantum numbers under the imposed symmetry for type II seesaw origin of neutrino mass and inelastic scalar DM.}
		\label{table2a}
	\end{table} 

\noindent
	{\bf Light Neutrino Masses}:  \\
The Yukawa Lagrangian relevant for the discussion is
	\begin{equation}
	-\mathcal{L} \supset Y_e \bar{L} H e_R + Y_{\nu} \bar{L^c} \Delta_L L +{\rm h.c.} + \mathcal{L}_{\rm DM}\,,
	\end{equation}
	where $\mathcal{L}_{\rm DM} $ is the Lagrangian for inelastic scalar or fermion DM as discussed below. 

\begin{figure}[h!]
		\centering
		\includegraphics[scale=0.60]{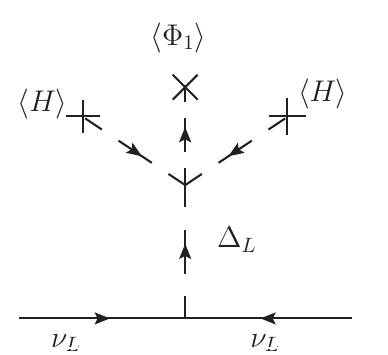}
		\caption{Type II Seesaw origin of neutrino mass.}
		\label{Fig0}
	\end{figure}

The relevant part of the scalar potential is given by:
	\begin{align}
	V & = -\mu^2_H H^{\dagger} H + \lambda_H (H^{\dagger}H)^2 +\mu^2_{\Delta} {\rm Tr}[\Delta^{\dagger}_L \Delta_L] + \lambda_{\Delta} {\rm Tr}[\Delta^{\dagger}_L \Delta_L]^2 +( \lambda_1 \Phi_1 H^T \Delta_L H +{\rm h.c.})
	\end{align}
We assume that $\mu_\Delta^2 > 0$ and $ \lambda_\Delta >0$. As a result $\Delta_L$ does not acquire any direct vev. However, the vev of 
$\Phi_1$ and $H$ can induce a small vev for $\Delta_L$ as:
\begin{equation}
	\langle \Delta^0_L \rangle \equiv v_L = -\frac{\lambda_1 \langle \Phi_1 \rangle v^2}{\mu^2_{\Delta}}.
	\end{equation}
As a result the light neutrino mass matrix is given by $m_\nu= Y_\nu v_L$.  The origin of light neutrino masses is shown in figure \ref{Fig0}. Note 
that in conventional type II seesaw, the induced vev is decided by trilinear term $\mu H^T \Delta_L H$ and hence for $\mu \sim \mu_{\Delta}$ it 
corresponds to a high scale seesaw like type I. However, here, the trilinear term is dynamically generated as $\mu =\lambda_1 \langle \Phi_1 \rangle$ 
via vev of the scalar $\Phi_1$. Note that $\Phi_1$ vev can be even lower than the electroweak scale. Thus for $\lambda_1 \langle \Phi_1 \rangle \ll 
\mu_{\Delta}$ (which can be achieved by suitable tuning of $\lambda_1$ and $\Phi_1$ vev), one can bring down the scale of type II seesaw $\mu_{\Delta}$ 
to a much lower scale.

Now we turn to comment on the viability of inelastic fermion and scalar DMs in this model.

\noindent
	{\bf Inelastic fermion dark matter}:  \\
The Lagrangian relevant for fermion DM is given by:
	\begin{equation}\label{fermion-term}
	\mathcal{L}_{\rm DM} =M \bar{\Psi} \Psi + \left (Y_L \frac{\overline{(\Psi_L)^c} \Psi_L \Phi_1 \Phi_2}{\Lambda}+Y_R \frac{\overline{(\Psi_R)^c} \Psi_R \Phi_1 \Phi_2}{\Lambda}+ {\rm h.c.} \right)
	\end{equation}
From Eq. \eqref{fermion-term}, we see that the dimension five terms generate 
small Majorana masses for $\Psi_L$ and $\Psi_R$  as $m_L= Y_L \langle \Phi_1\rangle \langle \Phi_2\rangle/\Lambda$ and $m_R= Y_R 
\langle \Phi_1\rangle \langle \Phi_2\rangle/\Lambda$ respectively. These Majorana masses split the Dirac fermoin $\Psi$ into two 
pseudo-Dirac states $\psi_1$ and $\psi_2$. The details of the corresponding DM Lagrangian remains same as discussed in the previous subsection.

\noindent
	{\bf Inelastic scalar dark matter}:  \\	
Similar to fermion DM realisation with type II seesaw, it is possible to have a scalar DM scenario as well. The corresponding particle content is 
shown in table \ref{table2a}. While the origin of neutrino mass remains the same as before, the relevant terms in the scalar DM Lagrangian can be 
written as follows.
\begin{equation}
\mathcal{L}_{\rm DM} \supseteq (D_\mu \eta)^\dagger (D^\mu \eta) - m^2_\eta \eta^\dagger \eta - (\lambda_2 \Phi_1 \Phi_2 \eta \eta + {\rm h.c.}) \,.
\label{scalar-term}
\end{equation}
	
From Eq. \eqref{scalar-term}, we see that the vevs of $\Phi_{1,2}$ generate a mass splitting between the real and imaginary parts of singlet 
scalar DM $\eta = (\eta_1+i\eta_2)/\sqrt{2}$. The corresponding mass splitting is given by: $\Delta m^2 = m^2_{\eta_2}-m^2_{\eta_1}= \lambda_2 
\langle \Phi_1 \rangle \langle \Phi_2 \rangle$. The mass splitting can be brought down to keV scale, necessary for explaining the anomalous XENON1T 
excess in section \ref{inelastic_dm}, by tuning $\lambda_2$ and $\Phi_1$ vev which does not play a role in $U(1)_X$ symmetry breaking. Further 
details of inelastic scalar dark matter remain same as discussed in the previous subsection.

\subsection{Radiative seesaw with inelastic DM}
Radiative seesaw has been one of the earliest proposals for low scale seesaw, see \cite{Cai:2017jrq} for a recent review. Due to additional 
loop suppressions and free parameters, natural realisation of low scale seesaw becomes possible in such frameworks. While there are many 
possible radiative seesaw, here we outline just one possibility that suits our desired phenomenology. In \cite{1807895}, a radiative seesaw 
model was introduced with the addition of a fermion doublet, a fermion singlet and a scalar singlet. Here we consider an alternate 
possibility with additional fermion singlet, scalar doublet and scalar singlet. Note that there are simpler realisation of one loop seesaw 
with dark matter, see for example~\cite{Ma:2006km}. However, the requirement of a sub-GeV inelastic scalar DM forces us to 
consider a complex scalar singlet into account because a scalar doublet with GeV scale components will be ruled out by precision data 
from Large Electron Positron (LEP) collider experiment \cite{Lundstrom:2008ai}.

\begin{table}
		\begin{tabular}{|c c|c|c|c|c|c|}
			\hline
			& &  $\eta$ & $\Phi$& $\Psi_{L,R}$& $ \chi$ \\ 
			\hline
			& $SU(2)_{L}$  & 1 & 1 & 1& 2\\
			\hline
			& $U(1)_{X}$  & 1 & 2 & 1& 1\\
			\hline
		\end{tabular}   
		\caption{New particles and their quantum numbers under the imposed symmetry for radiative seesaw model.}
		\label{table3}
	\end{table} 
	
The new particle content of the model is shown in table \ref{table3}. The relevant part of the Lagrangian consistent with $U(1)_X$ gauge 
symmetry is given by 
	
	\begin{figure}[h!]
		\centering
		\includegraphics[scale=0.60]{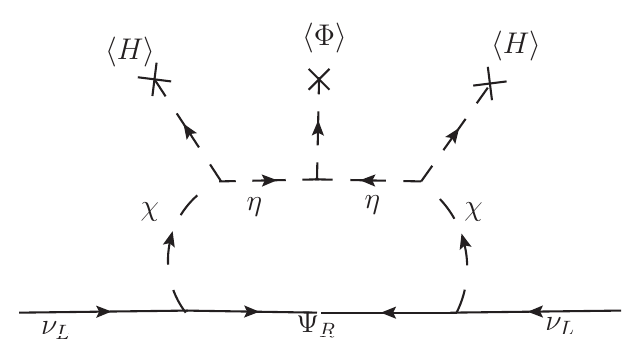}
		\caption{Radiative seesaw origin of light neutrino masses with dark sector particles in the loop.}
		\label{Fig0a}
	\end{figure}

	\begin{align}
	-\mathcal{L} \supset M \bar{\Psi} \Psi + \left (Y_{\nu} \bar{L} \tilde{\chi} \Psi_R + Y_L \Phi^{\dagger} \overline{(\Psi_L)^c} \Psi_L +Y_R \Phi^{\dagger} \overline{(\Psi_R)^c} \Psi_R + {\rm h.c.} \right)
	\end{align}
	The relevant part of the scalar potential is 
	\begin{align}\label{loop-potential}
	V \supset  m^2_{\eta} \eta^{\dagger} \eta + m^2_{\chi} \chi^{\dagger} \chi+ (\mu_1 \chi^{\dagger} H \eta + \mu_2 \eta \eta \Phi^{\dagger}+ {\rm h.c.} )
	\end{align}
	
The $U(1)_X$ symmetry is broken by a nonzero vev ($v_\phi$) of $\Phi$ to a remnant $Z_2$ symmetry under which $\Psi_{L, R}, \eta , \chi$ are odd 
while all other fields are even. As a result the lightest among $\Psi_{L, R}$, $\eta=\eta_1 + i \eta_2$ and $\chi=\chi_1 + i\chi_2$ can gives rise to a 
viable DM candidate. Since $\chi$ is a doublet it's mass can not be less than 45 GeV in order to avoid invisible Z-decay width. On the other hand, 
the mass of singlet fermion $\Psi$ and singlet scalar $\eta$ can be much smaller than the mass of $\chi$. In the effective theory, the vev of $\Phi$ 
generate small Majorana masses $m_L= Y_L v_\phi$ and $m_R=Y_R v_\phi$ for $\Psi_L$ and $\Psi_R$ respectively. As a result the Dirac fermion 
$\Psi$ splits up into two pseudo-Dirac states $\psi_1$ and $\psi_2$ with masses $M_1$ and $M_2$ respectively. Similarly the vev of $\Phi$ creates a mass 
splitting between the real and imaginary parts of $\eta$ through the term $\mu_2 \eta^2 \Phi + h.c.$ as given in the scalar potential \eqref{loop-potential}. 
The corresponding mass splitting is given by $\Delta m^2=m^2_{\eta_2} - m^2_{\eta_1}= 2\mu_2 v_{\phi}$. We will see that this mass splitting is related 
to the non-zero masses of light neutrinos in this model. 
 
The light neutrino mass arises at one loop level via the diagram shown in figure \ref{Fig0a}. The contribution to light neutrino mass 
can be estimated, in the mass insertion approximation \cite{Nomura:2017vzp}, to be 
	\begin{equation}
	(m_{\nu})_{ij} \simeq \frac{\mu^2_1 v^2 \mu_2 v_{\phi}}{(4\sqrt{2}) 16\pi^2} \frac{(Y_{\nu})_{ik} (M)_k (Y^T_{\nu})_{kj}}{M^6_{\chi}} 
        I_{\nu} (r_{\eta}, r_{k})\,, 
	\end{equation}
where $M_k$ is the mass of pseudo-Dirac states $\psi_k$ with $k=1,2$ and $v, v_{\phi}$ are vevs of neutral component of the SM Higgs doublet $H$ and 
the scalar singlet $\Phi$. The loop function $I_{\nu}$ is given by:
	\begin{equation}
	I_{\nu}(r_1, r_2) =\frac{1+r_1-2r_2}{2(1-r_1)^2 (1-r_2) (r_1-r_2)}-\frac{1}{2(1-r_1)^3 (r_1-r_2)^2} \bigg [ r_2+r_1(r_2-2r_1) \ln{r_1} + (1-r_1)^3 r_2 \ln{r_2} \bigg ]\,,
	\end{equation}
where the parameters $r_i$ are defined as $r_{\eta} = M^2_{\eta}/M^2_{\chi}$, and $r_{k} = M^2_{k}/M^2_{\chi}$ with $M^2_{\eta} = (m^2_{\eta_1} + m^2_{\eta_2})/2$ 
and  $M^2_{\chi} = (m^2_{\chi_1}+m^2_{\chi_2})/2$.
%
Similar to the seesaw models discussed earlier, here also light neutrino mass is proportional to the term in scalar potential which splits the scalar singlet ($\eta$) mass namely $\mu_2 \eta \eta \Phi^{\dagger}$. Non-zero $\mu_2$ implies non-zero mass splitting between scalar and pseudoscalar components of $\eta$. That is, 
$\Delta m^2= 2\mu_2 v_{\phi}$. Thus, we can rewrite the light neutrino mass formula as 
		\begin{equation}
	(m_{\nu})_{ij} \simeq -\frac{\mu^2_1 v^2 (m^2_{\eta_2} - m^2_{\eta_1})}{(8\sqrt{2}) 16\pi^2} \frac{(Y_{\nu})_{ik} (M)_k (Y^T_{\nu})_{kj}}{M^6_{\chi}} I_{\nu} (r_{\eta}, r_{k}) 
	\end{equation}
	If DM is dominantly from scalar singlet $\eta$, then the same mass splitting gives rise to inelastic nature of scalar DM. The details of which is same as discussed in the context of inverse seesaw and type II seesaw models. Similarly, one can have fermion singlet DM as well. The details of fermion DM Lagrangian 
remains same as discussed in previous models. Unlike inverse and type II seesaw, here DM mass or mass splitting gets directly related to the light neutrino mass 
due to the involvement of DM fields in the neutrino mass loop. Like in conventional radiative seesaw model, this model can also be realised as a low scale seesaw 
by suitable choices of parameters involved in the mass formula.

\section{Constraints on Low Scale $U(1)_X$}
\label{sec:constraint}
Since the neutrino and DM scenarios we discuss here are based on low scale $U(1)_X$ gauge symmetry, it is important to note the phenomenological constraints on such new physics scenario from available data. The $U(1)_X$ sector couples to the SM sector only via kinetic mixing of $U(1)_X$ and $U(1)_Y$ denoted by $\frac{\epsilon}{2} B^{\alpha \beta} Y_{\alpha \beta}$ in the Lagrangian where $B^{\alpha\beta}, Y_{\alpha \beta}$ are the field strength tensors of 
$U(1)_X, U(1)_Y$ respectively and $\epsilon$ is the kinetic mixing parameter. Even if we turn off such mixing at tree level, one can generate such mixing at one loop level since there are particles in the model which are charged under both $U(1)_Y$ and $U(1)_{X}$. Such one loop mixing can be approximated as $\epsilon \approx g_{1} g'/(16 \pi^2)$ \cite{Mambrini:2011dw} where $g_1, g'$ are gauge couplings of $U(1)_Y$ and $U(1)_{X}$ respectively. Thus, the relevant constraints will be applicable on effective DM-SM portal coupling $\epsilon g'$ and $Z'$ mass.

Such GeV scale gauge boson and couplings can be constrained from different low energy observations like neutrino trident production, rare kaon decay, 4 muon observations at BABAR experiment etc. As shown by the authors of \cite{Jho:2019cxq}, the current data allow the portal coupling $g' \lsim 0.001$ for $M_{Z'} \gsim 0.1$ GeV for muon-philic gauge boson. The portal coupling could be as large as even $0.005$ for $M_{Z'} \gsim 1.0$ GeV. As shown in \cite{Lindner:2018kjo}, the portal coupling can also be constrained from neutrino-electron scattering experiments like CHARM-II, GEMMA and TEXONO. Gauge boson mixing strength $\gsim 10^{-3}$ has been ruled out for gauge bosons of mass around the electroweak (EW) scale. For very light gauge bosons, this bound is even tighter $\mathcal{O}(10^{-6})$. LEP II data have put a lower bound on the ratio of new gauge boson mass to the new gauge coupling to be $M_{Z'}/g' \ge 7 $ TeV \cite{Carena:2004xs}. However, since we are interested in the low mass of the gauge boson, bounds from hadron colliders like ATLAS and CMS will not be very relevant. Similarly, LEP bound is also not applicable in such low mass regime. One can constrain $Z-Z'$ mixing and $Z'$ mass from electroweak precision measurements as well. However, for GeV scale $Z'$ mass with tiny kinetic mixing with SM $Z$ boson, such bounds do not apply \cite{Erler:2009jh}. For a detail of the direct search bounds on such a light gauge boson, one may refer to \cite{Deppisch:2019ldi}. Recently, a low scale $U(1)_X$ model was also studied in the context of flavour anomalies, dark matter and neutrino mass \cite{Borah:2020swo}.

\begin{figure}
\centering
\includegraphics[scale=0.6]{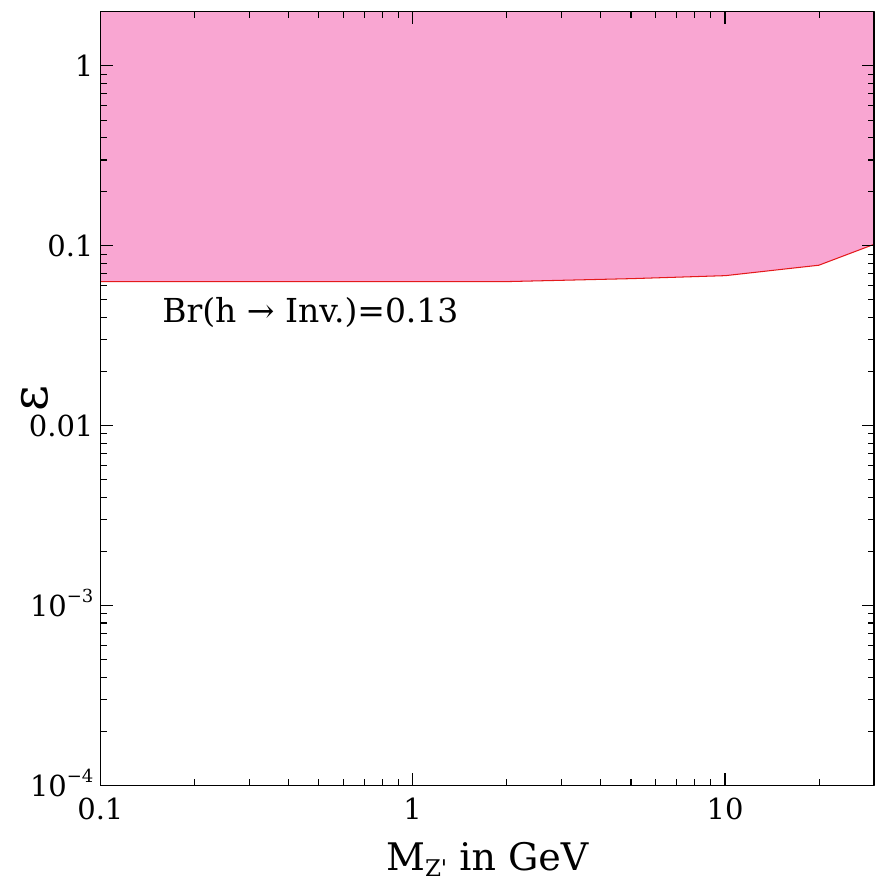}	
\caption{Constraints on the model parameters from Higgs invisible decay measurements. The pink shaded region is disallowed from the LHC limits on Higgs invisible decay width.}
\label{hinv}
\end{figure}

Apart from the $U(1)_X$ gauge sector, the additional scalar fields introduced for the purpose of spontaneous symmetry breaking, generation of light neutrino masses and inelastic DM are also constrained by experimental data. Though the singlet scalars do not directly couple to the SM particles, they can do so by virtue of their mixing with the SM Higgs. Precision electroweak measurements, perturbativity and unitarity of the theory as well as the LHC and LEP direct search \cite{Khachatryan:2015cwa,Strassler:2006ri} and measured Higgs signal strength constrains such mixing angle. In the scalar DM scenario, SM Higgs can directly couple to DM without relying on singlet-Higgs mixing. Since DM is in the GeV regime, such couplings will lead to SM Higgs decay into DM, contributing to its invisible decay width. The constraint on the Higgs invisible decay branching fraction from the ATLAS experiment at LHC is \cite{ATLAS:2020cjb}
	
	\begin{equation}
	\mathcal{B}(h \to \text{Invisible}) = \frac{\Gamma(h \to \text{Invisible})}{\Gamma(h \to SM) + \Gamma(h \to \text{Invisible}) } \leq 13\%.
	\end{equation}
Since all the scenarios discussed here contain light DM and $Z'$ gauge boson at or below GeV scale, SM Higgs can decay into them due to mixing of SM Higgs with singlet scalars and $Z-Z'$ mixing respectively. Since our analysis relies upon $Z-Z'$ mixing only, we consider Higgs decay to $Z' Z'$ and $Z Z'$ only. The corresponding decay widths are given by 
\begin{equation}
\Gamma (h \rightarrow Z' Z') = \frac{g^2 \epsilon^4}{128 \pi M^2_W} m^3_h \left ( 1-x-\frac{3}{4} x^2 \right) \sqrt{1-x}, \;\;\;\; x=\frac{4M^2_{Z'}}{m^2_h};
\end{equation}
\begin{equation}
\Gamma (h \rightarrow Z Z') = \frac{g^2 \epsilon^2}{128 \pi M^2_W} m^3_h \left ( 1-y-\frac{3}{4} y^2 \right) \sqrt{1-y}, \;\;\;\; y=\frac{(M_Z+M_{Z'})^2}{m^2_h}.
\end{equation}
Here $m_h$ denotes SM Higgs mass. Using these, we constrain the parameter space in $\epsilon-M_{Z'}$ plane using the bound mentioned above. The resulting parameter space ruled out from this bound is shown in figure \ref{hinv}. One can also constrain the model from the LHC measurements of SM Higgs decaying into light gauge bosons with four lepton final states \cite{Aaboud:2018fvk}. As will be clear from our final parameter space to be discussed later, such bounds are trivially satisfied for the region of parameter space we focus on.

	In some specific seesaw scenarios discussed above, electroweak multiplets are also introduced which couple directly to SM leptons. For example, the type II seesaw scenario considers a $SU(2)_L$ triplet which gives rise to a doubly charged physical scalar. Search for same sign dileptons at the LHC puts strict bound on such scalars, depending upon its branching ratio into specific leptonic final states \cite{Aaboud:2017qph}. Roughly, doubly charged scalar with masses below 800 GeV are currently disfavoured from such searches. Precision electroweak data can also rule out certain mass ranges of physical scalars belonging to this triplet \cite{Kanemura:2012rs}. The radiative seesaw model discussed above also contains an additional scalar doublet $\chi$ which remains inert (it does not acquire vev). The components of the doublet are also constrained from precision measurements as well as direct searches. For example, it strongly constrains the decay channel $Z \rightarrow \chi_1 \chi_2$ requiring $m_{\chi_1} + m_{\chi_2} > m_Z$. Here $\chi_{1,2}$ are real and imaginary parts of the neutral component of $\chi$, as defined before. Similar lower bound applies to the charged component of $\chi$ as well. Additionally, LEP precision data also rule out the region $m_{\chi_1} < 80 \; {\rm GeV}, m_{\chi_2} < 100 \; {\rm GeV}, m_{\chi_2} - m_{\chi_1} > 8 \;{\rm GeV}$ \cite{Lundstrom:2008ai}. Invisible Higgs decay constraints are applicable on neutral components in a way similar to the singlet scalar DM mentioned before.

As we show below, DM phenomenology in the GeV scale depends crucially on the gauge portal only and hence the bounds on scalar sector can be satisfied independently. In the gauge sector also, we find that for keV or smaller mass splitting between the two DM candidates, the constraints on lifetime of heavier DM dominates over all other constraints. For larger mass splitting where only one DM is present, the flavour bounds on gauge portal can become more relevant. Similarly, the bounds from observed neutrino mass squared differences and mixing can also be independently satisfied without affecting the phenomenology related to XENON1T excess and dark matter.

\section{Inelastic sub-GeV Dark Matter}
\label{inelastic_dm}
As discussed above, we extend the SM with a $U(1)_X$ symmetry which can simultaneously accommodate non-zero neutrino mass and inelastic dark matter, comprising of two quasi-degenerate components $\chi_1$ and $\chi_2$ with masses $M_1$ and $M_2$ respectively. We assume a small mass splitting $\Delta m = M_2-M_1$, where $\Delta m << M_{1,2}$, between the two DM components and this mass splitting is non-trivially related to non-zero light neutrino masses such that degenerate DM components will lead to vanishing light neutrino masses.

The inelastic nature of sub-GeV DM is particularly appealing in the light of recent experimental results from XENON1T \cite{Aprile:2020tmw}. The inelastic down scattering of a sub-GeV scale DM with the 
electrons in Xenon atoms provides a viable explanation for XENON1T excess of electron recoil events near 1-3 keV energy~\cite{Harigaya:2020ckz, Borah:2020jzi}. In this case, a heavier DM inelastically scatters off an electron and gets converted to the lighter DM component. The small mass splitting of keV scale between the two DM components $\Delta m$ is transferred to the electron recoil energy. Due to such tiny splitting, the heavier DM is long lived and as a result the total relic density of DM is the sum of individual contributions. At this juncture 
we note that in the original inelastic DM~\cite{TuckerSmith:2001hy} proposal the lighter DM component 
scatters off a nucleon and gets converted to a heavier DM component. If the mass splitting between the two 
components is much larger than the nuclear recoil energy then such processes are forbidden. In particular, 
the SM Z-boson mediated interactions in direct search experiments can be forbidden if the DM is 
inelastic~\cite{Arina:2011cu,Arina:2012aj,Arina:2012fb,Cui:2009xq, Okada:2019sbb}. The main difference between the two scenarios is that in the former case the life time of heavier component is required to be longer than the age of the universe, while 
in the latter it is not required. We will discuss the corresponding results for the latter case as well where DM in the present universe is effectively in terms of the lighter component only as the heavier component can decay in early epochs due to large mass splitting.
 
The relic abundance of two component DM can be found by numerically solving the corresponding Boltzmann 
equations. Let $n_2$ and $n_1$ are the total number densities of two dark matter candidates $\chi_{2}$ and $\chi_1$ 
respectively. The two coupled Boltzmann equations in terms of $n_2$ and $n_1$ are given below,   
\begin{widetext}
\begin{eqnarray}
\frac{dn_{2}}{dt} + 3n_{2} H &=& 
-\langle{\sigma {\rm{v}}}_{\chi_2 \chi_2 \rightarrow {X \bar{X}}}\rangle 
\left(n_{2}^2 -(n_{2}^{\rm eq})^2\right)
- {\langle{\sigma {\rm{v}}}_{\chi_2 \chi_2
\rightarrow \chi_1 \chi_1}\rangle} \bigg(n_{2}^2 - 
\frac{(n_{2}^{\rm eq})^2}{(n_{1}^{\rm eq})^2}n_{1}^2\bigg) \nonumber \\&-&  \langle{\sigma {\rm{v}}}_{\chi_2 \chi_1 \rightarrow {X \bar{X}}}\rangle 
\left(n_{1} n_2 -n_{1}^{\rm eq} n_{2}^{\rm eq}  \right)\,, \nonumber 
%
\label{boltz-eq1} \\
\frac{dn_{1}}{dt} + 3n_{1} H &=& -\langle{\sigma {\rm{v}}}
_{\chi_1 \chi_1 \rightarrow {X \bar{X}}}\rangle \left(n_{1}^2 -
(n_{1}^{\rm eq})^2\right) 
+ {\langle{\sigma {\rm{v}}}_{\chi_2 \chi_2 \rightarrow {\chi_1} \chi_1}\rangle} 
\bigg(n_{2}^2 - \frac{(n_{2}^{\rm eq})^2}{(n_{1}^{\rm eq})^2}
n_{1}^2\bigg) \nonumber \\&-& \langle{\sigma {\rm{v}}}_{\chi_2 \chi_1 \rightarrow {X \bar{X}}}\rangle 
\left(n_{1} n_2 -n_{1}^{\rm eq} n_{2}^{\rm eq}  \right) \ \nonumber \\
\label{boltz-eq2} 
\end{eqnarray}
\end{widetext}
where, $n^{\rm eq}_i$ is the equilibrium number density of dark matter species $i$ and $H$ denotes the Hubble 
expansion parameter. The thermally averaged annihilation and coannihilation processes ($\chi_i \chi_j 
\rightarrow X \bar{X}$) are denoted by $\langle{\sigma {\rm{v}}} \rangle$, where X denotes all particles to 
which DM can annihilate into. Since we consider GeV scale DM, the only annihilations into light SM fermions 
can occur, such as $e^-,\mu^-, \nu_e,\nu_\mu,\nu_\tau, u,d,s$. The only available channel for annihilation 
of $\chi_{1,2}$ to light SM fermions is through $Z-Z'$ mixing. Additionally small mass splitting between the 
two DM components lead to efficient coannihilations while keeping their conversions into each other sub-dominant. 
We have solved these two coupled Boltzmann equations using \texttt{micrOMEGAs} \cite{Belanger:2014vza}. Due to 
tiny mass splitting, we find almost identical relic abundance of two DM candidates. Thus each of them constitutes 
approximately half of total DM relic abundance in the universe, {\it i.e.} $n_2 \approx n_1 \approx n_{\rm DM}/2$. 
We then constrain the model parameters by comparing with Planck 2018 limit on total DM abundance $\Omega_{\text{DM}} 
h^2 = 0.120\pm 0.001$ \cite{Aghanim:2018eyx}. Here $\Omega_{\rm DM}$ is the density parameter of DM and 
$h = \text{Hubble Parameter}/(100 \;\text{km} ~\text{s}^{-1} \text{Mpc}^{-1})$ is a dimensionless parameter of 
order one.

As discussed above, we assume $\chi_2$ is heavier than $\chi_1$ with a small mass splitting $\Delta m = M_2-M_1$ 
between the two components. Moreover, we assume $\Delta m$ of keV scale in order to explain the XENON1T anomaly. 
For a fixed incoming velocity $v$ of $\chi_2$, the differential scattering cross section for $\chi_2 e \rightarrow 
\chi_1 e$ can be given as

\begin{equation}
	\frac{d \langle \sigma v \rangle }{d E_r} = \frac{\sigma_e}{2 m_e }\int^{v_{esc}}_0 dv \frac{f(v)}{v} \int_{q-}^{q+}dq~~ a^2_0 q |F(q)|^2 K(E_r,q)\,,
	\label{Event}
\end{equation}
where $m_e$ is the electron mass, $\sigma_e$ is the free electron cross section at fixed momentum transfer 
$q=1/a_0$, where $a_0 = \frac{1}{\alpha m_e}$ is the Bohr radius with $\alpha = \frac{e^2}{4 \pi}=\frac{1}{137}$ being 
the fine structure constant, $E_r$ is the recoil energy of electron and $K(E_r, q)$ is the atomic excitation factor. We assume the DM form factor to be unity. In 
this paper the atomic excitation factor is adopted from \cite{Roberts:2019chv}. For $E_r=(1-5)$ keV, the scattering happens dominantly with electrons in the 3s shell. The Atomic excitation factor $K(E_r,q)$ is independent of $E$ before it reaches the threshold of the next quantum energy level. Since most of the signal events have a recoil energy in a range $2-3$ keV \cite{Aprile:2020tmw}, so one can use $K(E_r,q) \simeq K(\Delta m, q) \simeq K(2\; {\rm keV}, q)$ \cite{Roberts:2019chv} for the calculation.  
In the above equation \eqref{Event}, $f(v)$ is the local DM velocity distribution function which can always be normalised to unity i.e. $\int f(v) dv=1$. $f(v)$ can be taken as a pseudo-Maxwellian distribution given by
	\begin{equation}
		f(v)= A v^2 {\rm Exp}[-(v-v_m)^2/2\sigma^2_v]
	\end{equation}
	where $A$ is the normalisation constant, $v_m$ is the average velocity which we consider to be $v_m=1\times10^{-3}$ and $\sigma_v$ is the DM velocity dispersion.
The free electron scattering cross-section in this case is given by:
\begin{equation}
\sigma_e = \frac{16 \pi \alpha_Z \alpha^{'} \epsilon^2 m^2_e  }{M^4_{Z'}}
\label{DM-electron-scattering}
\end{equation}
where $\alpha_Z=\frac{g^2}{4 \pi}$, $\alpha^{'}=\frac{g'^2}{4 \pi}$ and $\epsilon$ is the kinetic mixing parameter 
between $Z$ and $Z'$ mentioned earlier which we take to be $\epsilon \leq 10^{-3}$. It should be noted that $\sigma_e$ 
is independent of DM mass as the reduced mass of DM-electron is almost equal to electron mass for GeV scale DM mass we are considering. 

Unlike the elastic case, the limits of integration in Eq.~\eqref{Event} are determined depending on the relative 
values of recoil energy ($E_r$) and the mass splitting between the two DM components. 
	
	For $E_r \geq \Delta m$
	\begin{equation}
	q_\pm=M_{2} v \pm \sqrt{M^2_{2} v^2 -2M_{2}(E_r-\Delta m)}\,.
	\end{equation}
	And for $E_r \leq \Delta m$
	\begin{equation}
	q_\pm=\sqrt{M^2_{2} v^2 -2M_{2}(E_r-\Delta m)} \pm M_{2} v \,.	
	\end{equation}
	
\begin{figure}
\centering
\includegraphics[scale=0.45]{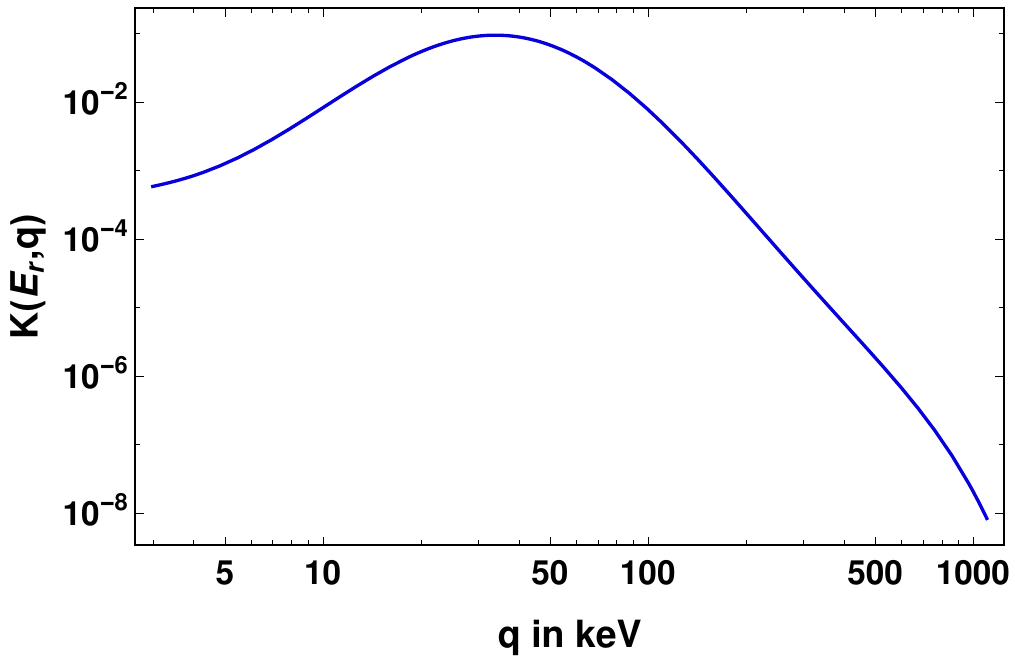}	
\hfil
\includegraphics[scale=0.6]{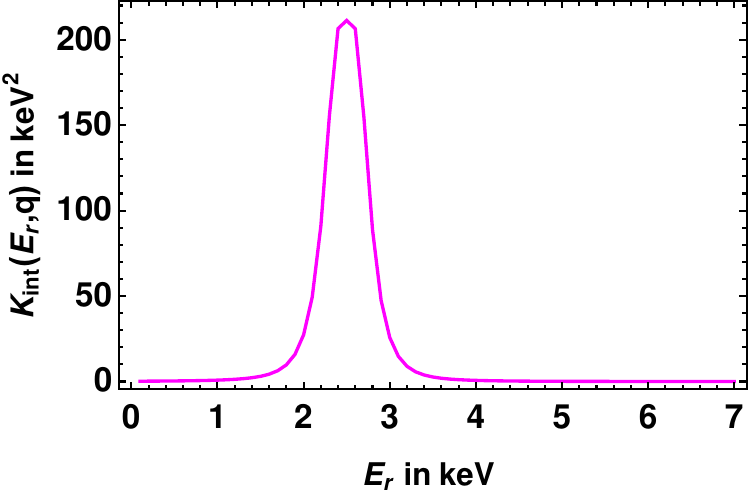}
\caption{Left panel: Dependence of Atomic excitation factor on momentum transferred. Right panel: The atomic excitation factor, after the q integration, is plotted as a function of the transferred recoil energy $E_r$.  }
\label{aef}
\end{figure}

The dependency of atomic excitation factor on the momentum transferred $q$ is shown in figure~\ref{aef}. Here the dominant contribution comes from the bound states with principal quantum number $n=3$ as their binding energy is around a few 
keVs. In the right panel of figure~\ref{aef}, we have shown the plot for the integration of momentum transferred times the 
atomic excitation factor $\big({\it i.e.}K_{int}(E_r,q)=\int_{q-}^{q+} q dq K(E_r,q)\big)$ as a function of the recoil energy 
$E_r$ for $M_1=0.3$GeV and $\Delta m = 2$keV. The figure shows a peak around $E_r \simeq \Delta m$ since the $q_{-}$ 
approaches to zero and the momentum transfer maximising this factor is available. It is worth mentioning that such kind 
of enhancement is a characteristic feature of inelastic scattering. 
	
The differential event rate for the inelastic DM scattering with electrons in Xenon atom, {\it i.e} $\chi_2 e 
\rightarrow \chi_1 e$, can be given as: 
\begin{equation}
\frac{dR}{dE_r}=n_T n_{\rm DM} \frac{d \sigma v}{d E_r}
\label{event_rate}
\end{equation}
\begin{figure}
\centering
\includegraphics[scale=0.4]{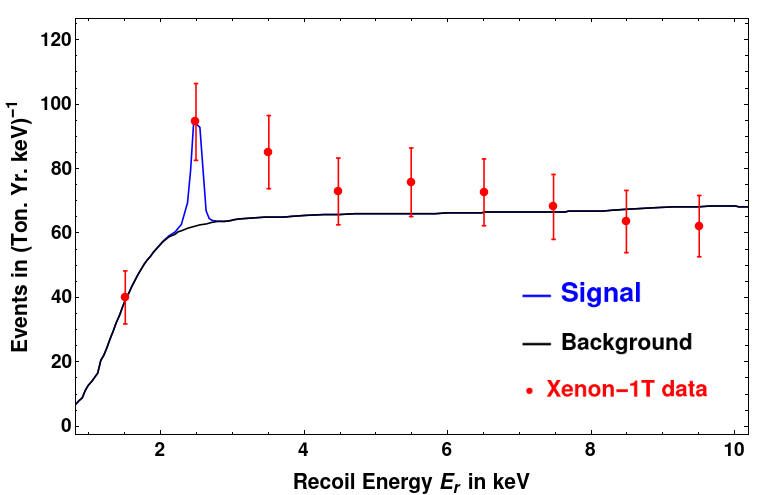}	
\includegraphics[scale=0.34]{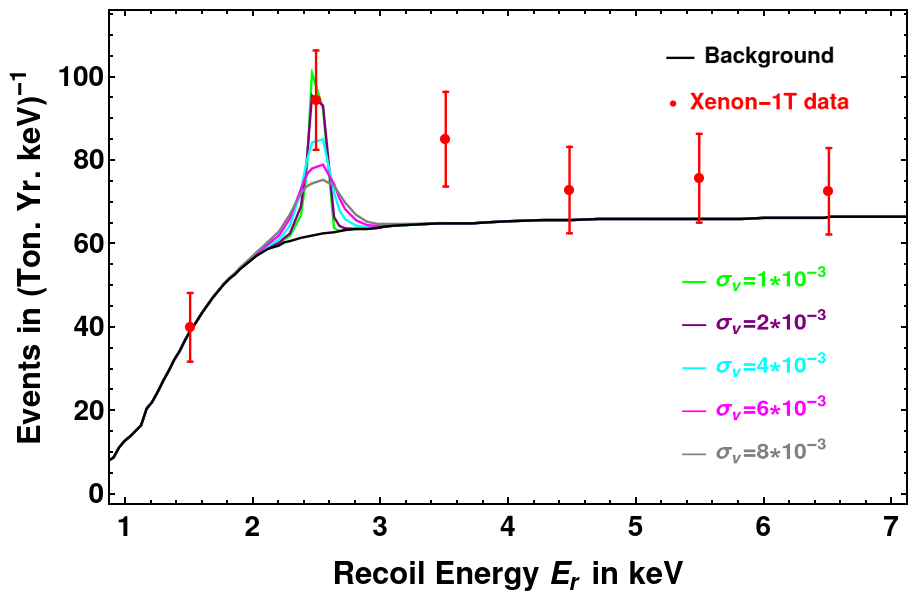}
\caption{Fit to XENON1T data with inelastic DM in our model}
\label{xenonfit}
\end{figure}
where $n_T=4\times10^{27}$ $ {\rm Ton}^{-1}$ is the number density of Xenon atoms and $n_{DM}$ is the number density of the dark matter particle.

The detected recoil energy spectrum can be obtained by convolving the above equation~\eqref{event_rate} with the energy resolution of the detector. Incorporating the detector efficiency, the energy resolution of the detector is given by a Gaussian distribution with an energy dependent width, 
	\begin{equation}
		\zeta(E,E_r)=\frac{1}{\sqrt{2 \pi \sigma^2_{det}}}{\rm Exp}\Big[-\frac{(E-E_r)^2}{2 \sigma^2_{\rm det}}\Big] \times \gamma(E)
	\end{equation}
	where $\gamma(E)$ is the detector efficiency which is reported in figure~2 of \cite{Aprile:2020tmw} and the width $\sigma_{\rm det}$ is given by 
	\begin{equation}
		\sigma_{\rm det}(E)= a \sqrt{E} + b E
	\end{equation} 
	where $a=0.3171$ and $b=0.0037$.
	Thus the final detected recoil energy spectrum is given by
	\begin{equation}
		\frac{dR_{\rm det}}{dE_r}=\frac{n_T n_{\rm DM} \sigma_e a^2_0}{2 m_e} \int dE ~~\zeta(E,E_r) \Bigg[\int dv \frac{f(v)}{v} \int_{q-}^{q+} dq~~ q K(E_r,q)\Bigg] 	
	\end{equation}

\subsection{Fermion DM}
For details of the inelastic fermion DM Lagrangian relevant for DM phenomenology, please refer to the subsection \ref{subsec:iss}. The fit to XENON1T data of electron recoil excess in an inelastic down scattering of DM scenario of our model is shown in figure~\ref{xenonfit}. To obtain such a fit, the mass splitting is taken to be $\Delta m = 2.5$ keV while heavier DM mass is taken to be 0.3 GeV. DM velocity is taken to be $v \simeq 1 \times 10^{-3}$ which is consistent with non-relativistic nature of CDM. The electron recoil cross-section corresponding to this obtained fit is $\sigma_e =1.2 \times 10^{-17}$GeV$^{-2} $.  In the bottom panel of Figure~\ref{xenonfit}, we have shown the fit considering different velocity dispersion for the DM particle. Clearly as we increase the velocity dispersion the peak in the spectrum giving an appreciable fit flattens out and the fit no longer touches the XENON1T signal within $E_r=2-3$keV for larger $\sigma_v$.  


\begin{figure}[h!]
	\centering
	\includegraphics[scale=0.65]{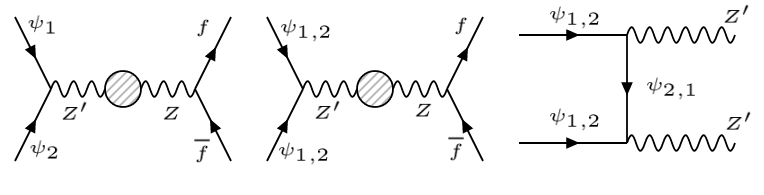}
	\caption{Dominant channels for relic abundance of fermionic DM}
	\label{fermion_relic_feyn}
\end{figure}

\begin{figure}[h!]
	\centering
	\includegraphics[scale=0.35]{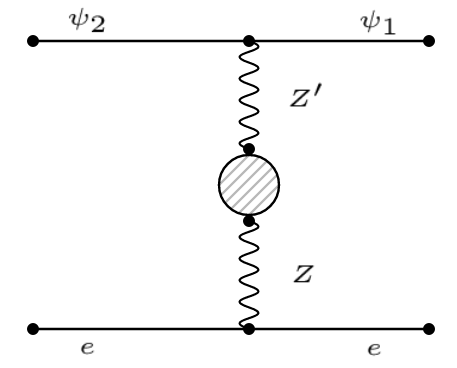}
%
	\hfil
	\includegraphics[scale=0.35]{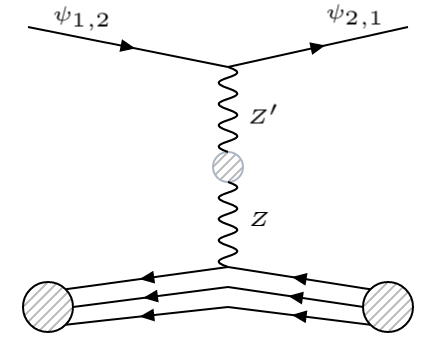}
	\caption{Left panel: Inelastic down-scattering of the heavier DM particle $\psi_2$ off
		the electron e into the lighter particle $\psi_1$, mediated by the $U(1)_X$ gauge boson $Z'$ that mixes kinetically with SM $Z$ boson. Right panel: Inelastic scattering of fermion DM off
		a nucleon, mediated by the $U(1)_X$ gauge boson $Z'$ that mixes kinetically with SM $Z$ boson.}
		\label{fermionDM_scatter}
\end{figure}
\begin{figure}[h!]
	\includegraphics[scale=0.5]{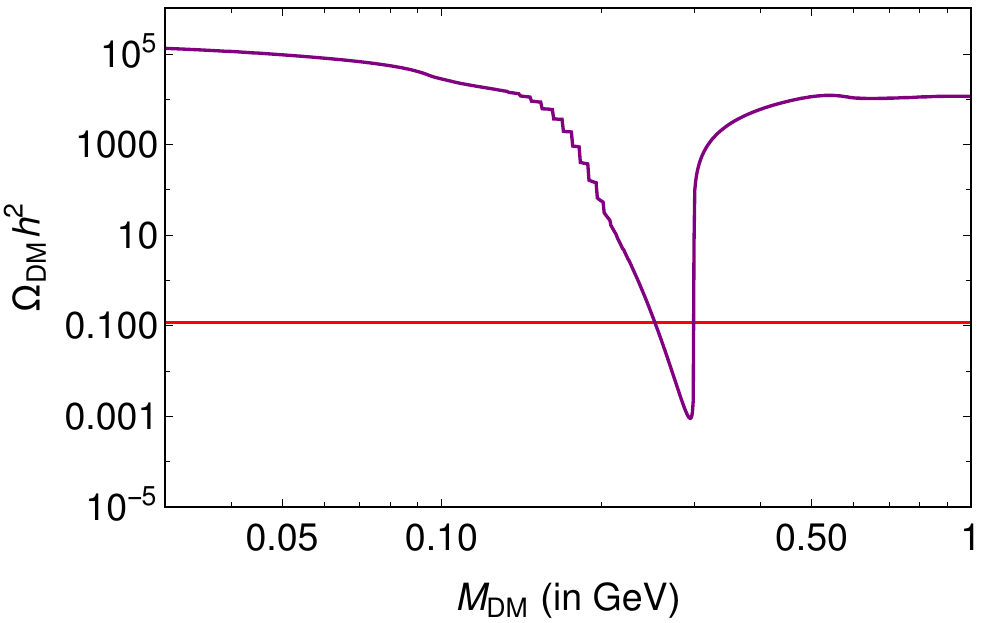}
	\caption{Relic density vs mass of inelastic fermion DM $M_{\rm DM} =M_1 \approx M_2$ where the resonance corresponds to $Z'$ boson with mass $0.6$ GeV .}
		\label{fermionDM}
\end{figure}

\begin{figure}[h!]
	\centering
	\includegraphics[scale=0.7]{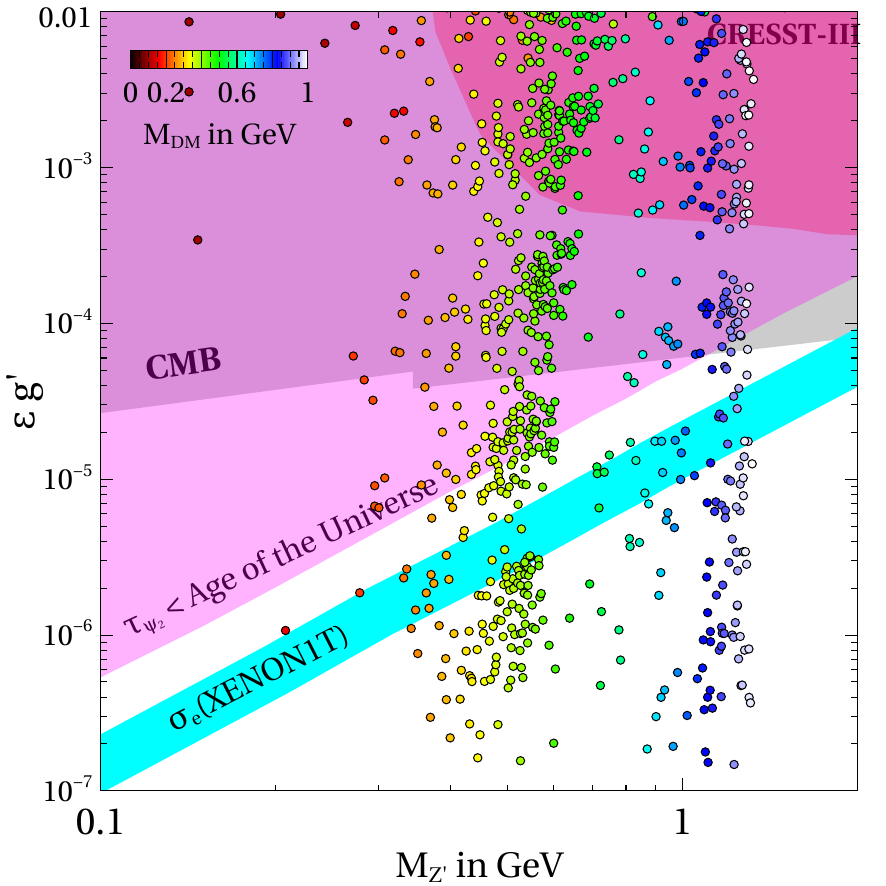}
	\caption{Summary plot for fermionic inelastic DM showing the final parameter space from various relevant constraints.}
	\label{fermionsummary}
\end{figure}

Relic abundance of fermion DM is mainly governed by the annihilation and coannihilation diagrams shown in figure \ref{fermion_relic_feyn}. Since among the newly introduced particles, only DM and $Z'$ gauge bosons are kept near the GeV regime, the final state particles can have either $Z'$ or light SM fermions. For tiny mass splitting of keV or below, both the DM components can be stable. There can be a conversion process also where heavier DM converts into the lighter one. However, for tiny mass splitting conversion is not very efficient, as we will see shortly. The same $Z-Z'$ portal can also give rise to inelastic DM-electron and inelastic DM-nucleus scattering cross section as shown in figure \ref{fermionDM_scatter}. While we show only the down-scattering of DM with electron (as favoured from XENON1T point of view), the DM-nucleon scattering can be both up or down type depending upon the mass splitting. While DM-electron scattering cross section is given by equation \eqref{DM-electron-scattering}, the spin-independent DM-nucleus scattering cross section is given by
\begin{equation}
\sigma^{SI}_{N \chi} = \frac{\mu^2_{ N \chi}}{\pi} \frac{g^2 g'^2 \epsilon^2}{M^4_{Z'}} \frac{[Z f_p + (A-Z)f_n]^{2}}{A^2}
\label{DM-nucleon-scattering}	
\end{equation} 
where $\chi$ denotes the DM particle, A and Z are the mass number and atomic number of the target nucleus respectively and $\mu_{ N \chi}=\frac{m_N m_\chi}{m_N + m_\chi}$ is the reduced mass. $f_p$ and $f_N$ are the interaction strengths for proton and neutron respectively. The occurrence of this process solely depends on the mass splitting between the two states. In fact, the minimum velocity of the DM needed to register a recoil inside the detector is
given
\begin{equation}
v_{\rm min}= c \sqrt{\frac{1}{2 m_N E_r}}\left({\frac{m_n E_r}{\mu_{N\chi}}+\Delta m}\right)	
\end{equation}
If the mass splitting is above a few hundred keV, then the inelastic scattering will be forbidden. On the other hand, the elastic DM-nucleon scattering is much smaller due to velocity suppression.

Variation of relic abundance of fermion DM as a function of its mass is shown in figure \ref{fermionDM} for a set of fixed benchmark parameters. Clearly, due to tiny mass splitting (2.5 keV) between two DM candidates and identical gauge interactions, their relic abundances are almost identical. The DM annihilation due to s-channel mediation of $Z'$ gauge boson is clearly visible from this figure where correct relic of DM is satisfied near the resonance region $M_{\rm DM} \approx M_{Z'}/2$. Final summary plot for fermion DM with 2.5 keV mass splitting is shown in figure \ref{fermionsummary} in the plane of $\epsilon g'$ vs $M_{Z'}$ since $ \epsilon g'$ is the 
relevant parameter for the portal linking dark sector and SM. For performing a random scan for the relic abundance of such two component DM, we fixed the mass splitting at $\Delta m= 2$keV. The gauge 
coupling was varied in $\mathcal{O}(1)$ i.e. in the interval $(1, 3.54)$ while the kinetic mixing parameter was varied in the range $(10^{-7}, 10^{-2})$. We also varied continuously DM mass from $(0.1-1)$ GeV and $Z'$ boson mass from $(0.1-2)$ GeV. Clearly the points residing in the cyan coloured solid band are satisfied by all relevant constraints. 


Since the mass splitting between $\psi_1$ and $\psi_2$ is kept at keV scale $\Delta m= \mathcal{O}(keV)$, there can be decay modes
like $\psi_2 \rightarrow \psi_1 \nu \overline{\nu}$  mediated by $Z-Z'$ mixing. If both the DM components are to be there in the present 
universe, this lifetime has to be more than the age of the universe that is $\tau_{\psi_2} > \tau_{\rm Univ.}$. The decay width of this 
process is $\Gamma(\psi_2 \rightarrow \psi_1 \nu \overline{\nu})= \frac{g^2 g'^2 \epsilon^2 (\Delta m)^5}{160 \pi^3 M^4_{Z'}}$. Thus, imposing 
the lifetime constraint on heavier DM, we show the excluded parameter space by the magenta colored region. 

We also show the parameter space excluded by 
the recent results from CRESST-III on low mass DMs. The solid band of cyan colour corresponds to free electron cross section $\sigma_e=10^{-17}-10^{-16}$ 
GeV$^{-2}$ which is required to obtain the fit for the XENON1T excess for a DM of mass around 1 GeV with a typical DM velocity of order $\mathcal{O}(10^{-3})$. 
The points satisfied by the observed relic density constraint of DM are shown by the coloured scattered points where the colour coding gives the information of the DM mass. As mentioned earlier, the other bounds like the ones from flavour physics experiments are weaker compared to the ones shown in the summary plot \ref{fermionsummary}. The bounds from dark photon searches at BABAR \cite{Lees:2014xha} will lead to an exclusion line in the range $\epsilon g' \sim 10^{-4}-10^{-3}$ which remains weaker than the lifetime bound and hence not shown.

We also incorporate the Planck bound on DM annihilation into charged leptons \cite{Aghanim:2018eyx}, specially $ e^- e^+, \mu^- \mu^+ $ pairs which are dominant in the region of our interest. The Cosmic microwave background (CMB) anisotropies, very precisely measured by the Planck experiment, are sensitive to energy injection in the intergalactic medium (IGM) from DM annihilations. The effective parameter constrained by CMB anisotropies is 
\begin{equation}
P_{\rm ann} = f_{\rm eff} (z) \frac{\langle \sigma v \rangle }{M_{\rm DM}} 
\end{equation}
where $f_{\rm eff}(z)$ is the efficiency factor characterising the fraction of energy transferred by the DM annihilation processes into the IGM. While the efficiency factor is redshift dependent, CMB anisotropies are most sensitive to redshift $z \sim 600$. For earlier works on imprint of DM annihilations on CMB please see \cite{Padmanabhan:2005es, Slatyer:2009yq, Slatyer:2012yq, Slatyer:2015jla}. Considering the efficiency factor to be close to unity for DM mass ranges of our interest, the Planck 2018 bound $P_{\rm ann} < 3 \times 10^{-11} {\rm GeV}^{-3}$ at $95\%$ C.L. can be used to constrain the model parameters. We consider the dominant s-wave coannihilations of DM into charged leptons and assume $3 M_{\rm DM} = M_{Z'}$ to derive the CMB exclusion line in figure \ref{fermionsummary}. Since DM relic is satisfied mostly around the $Z'$ resonance, this relation will not change much for other allowed points as well. As can be seen from figure \ref{fermionsummary}, the CMB bound on DM annihilation can be even stronger than the lifetime bound for DM mass beyond 1 GeV. It however leaves plenty of parameter space consistent with DM relic as well as XENON1T favoured scattering cross section of DM with electrons.

The dominance of $Z'$ resonance in DM annihilation is visible from the scattered points in figure \ref{fermionsummary}. It can be seen that DM with a particular mass around $M_{Z'}/2$ can satisfy relic independent of portal coupling $\epsilon g'$. This is due to the resonance feature around $M_{\rm DM} \approx M_{Z'}/2$ which was also noticed in figure \ref{fermion_relic_feyn}.

\subsection{Scalar DM}
We now turn to find a viable inelastic scalar DM in a scenario where the SM is augmented with a $U(1)_X$ symmetry. For details of the inelastic scalar DM Lagrangian relevant for DM phenomenology, please refer to the subsection \ref{subsec:iss}.
 
\begin{figure}[h!]
	\includegraphics[scale=0.7]{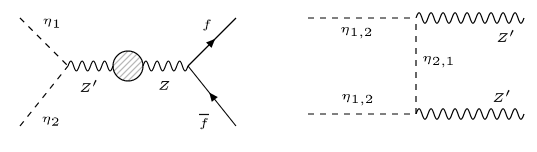}
	\caption{Dominant channels for relic abundance of scalar DM}
	\label{scalar_relic_feyn}
\end{figure}

\begin{figure}[h!]
	\includegraphics[scale=0.35]{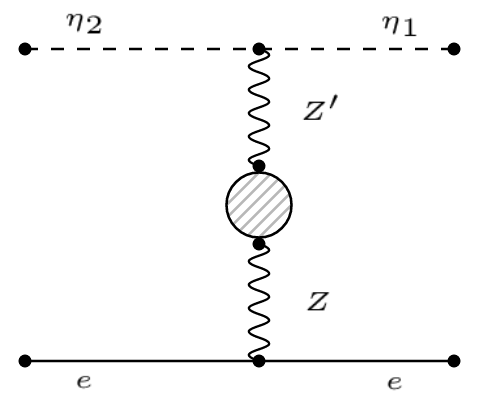}
%
	\hfil
	\includegraphics[scale=0.40]{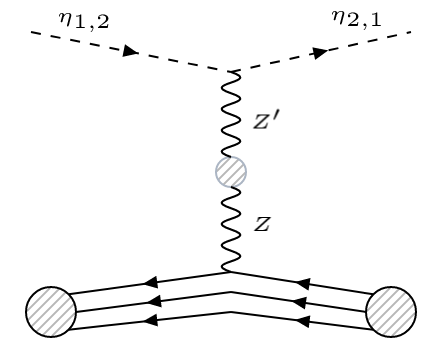}
	\caption{Left panel: Inelastic down-scattering of the heavier scalar DM particle $\eta_2$ off
		the electron e into the lighter particle $\eta_1$, mediated by the $U(1)_X$ gauge boson $Z'$ that mixes kinetically with SM $Z$ boson. Right panel: Inelastic scattering of scalar DM off
		a nucleon, mediated by the $U(1)_X$ gauge boson $Z'$ that mixes kinetically with SM $Z$ boson.}
\label{scalarDM}
\end{figure}

Considering only scalar DM and $Z'$ to be light and in the GeV regime among the newly introduced particles, the dominant annihilation and coannihilation channels which control scalar DM relic abundance are shown in figure \ref{scalar_relic_feyn}. The diagrams with light SM fermions or $Z'$ in final states dominate. Due to tiny mass splitting, the conversion from heavier to lighter DM is not very efficient, as noted in the discussion of fermion DM before. These same interactions which govern the annihilation processes can also lead to DM-electron and DM-nucleus scattering. The heavier DM particle $\eta_2$ (say) inelastically scatters off an electron in the Xenon atom in the detector 
and gets converted to the lighter state $\eta_1$ (say), as demonstrated in the left panel plot of figure~\ref{scalarDM}. The small mass difference $\Delta m$ is converted into the electron recoil energy, which may explain the excess of events observed by the XENON1T experiment. Similarly, the right panel plot of figure~\ref{scalarDM} shows the inelastic DM-nucleus scattering which can be both up or down type depending upon the mass splitting. The corresponding scattering cross sections can be found in a way similar to the fermion DM scenario. Note that, unlike the fermion case, there is no elastic scattering diagram of scalar DM mediated by $Z'$.


\begin{figure}[h!]
	\includegraphics[scale=0.5]{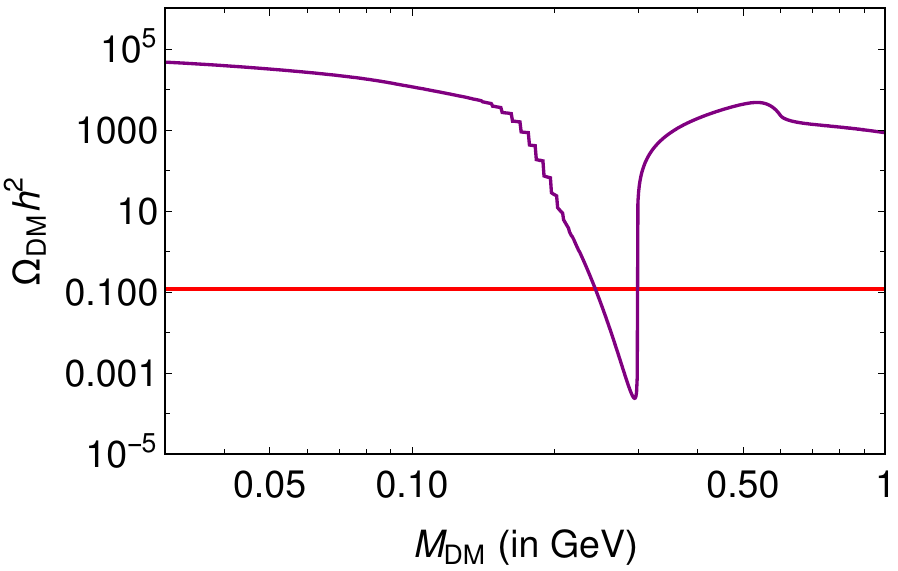}
	\caption{Relic density vs mass of inelastic scalar DM $M_{\rm DM} =M_1 \approx M_2$ where the resonance corresponds to $Z'$ boson with mass $0.6$ GeV .}
\label{scala_Dm_fig}
\end{figure}	

\begin{figure}[h!]
	\centering
	\includegraphics[scale=0.7]{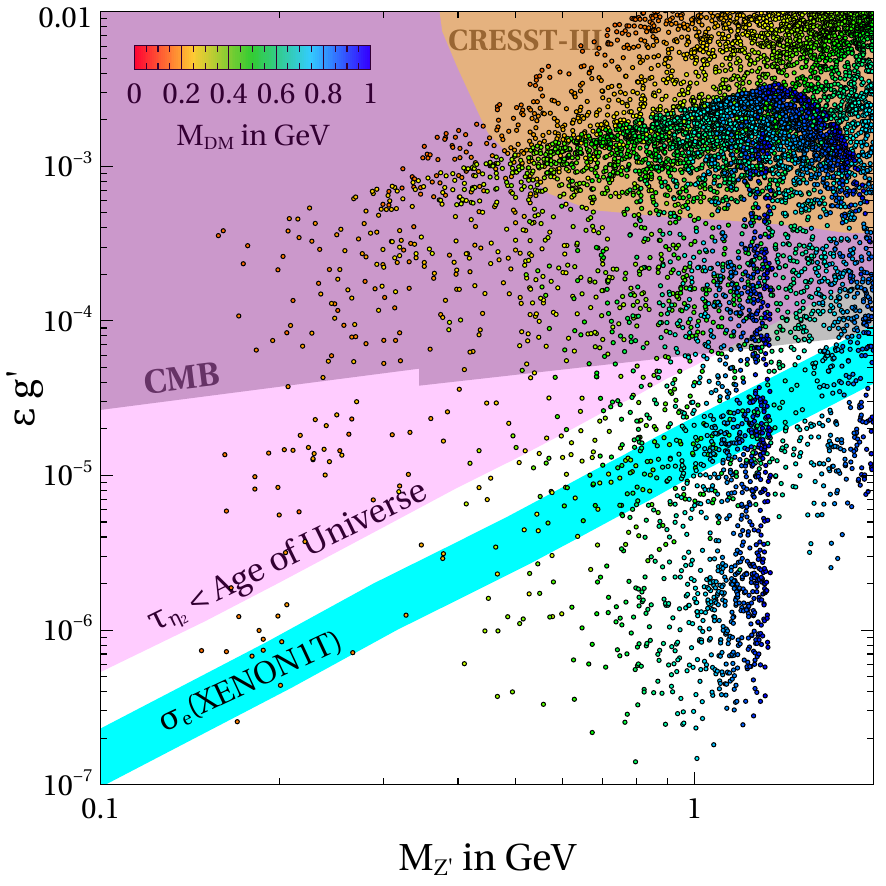}
	\caption{Summary plot for inelastic scalar DM showing the final parameter space from various relevant constraints.}
	\label{scalarsummary}
\end{figure}	

Similar to fermion DM, scalar DM relic in the sub-GeV regime also depends crucially on $Z'$ mediated coannihilation channel. The variation of its relic abundance as a function of DM mass is shown in figure \ref{scala_Dm_fig} for a fixed set of benchmark. Clearly, the $Z'$ resonance is visible there. Similar to the fermion case, we show in figure~\ref{scalarsummary} the final parameter space in the plane of $\epsilon g'$ vs $M_{Z'}$ allowed by all the relevant constraints. The summary plot looks very similar to the one for fermion and can be explained in a similar way. Similar to the fermion DM case here also the heavier DM can decay into lighter DM and two neutrinos at tree level. The corresponding lifetime bound is shown in figure \ref{scalarsummary}.

\subsection{DM conversion at late epochs}
Although relative abundance of the two DM candidates $\chi_1$ and $\chi_2$ are expected to be approximately the half of total DM relic 
abundance from the above analysis based on chemical decoupling of DM from the SM bath, there can be internal conversion happening between the two DM candidates via processes like $\chi_2 \chi_2 \rightarrow \chi_1 \chi_1, \chi_2 e \rightarrow \chi_1 e$ until later epochs. While such processes keep the total DM density conserved, they can certainly change the relative proportion of two DM densities. It was pointed out by \cite{Harigaya:2020ckz, Baryakhtar:2020rwy} as well as several earlier works including \cite{Finkbeiner:2007kk, Batell:2009vb}. In these works, DM is part of a hidden sector comprising a gauged $U(1)_X$ which couples to the SM particles only via kinetic mixing of $U(1)_X$ and $U(1)_Y$, denoted by $\epsilon$, same as our scenario. Thus, although the DM-SM interaction is suppressed by $\epsilon^2$ leading to departure from chemical equilibrium at early epochs, the internal DM conversions like $\chi_2 \chi_2 \rightarrow \chi_1 \chi_1$ can happen purely via $U(1)_X$ interactions and can be operative even at temperatures lower than chemical freeze-out temperature, specially when $U(1)_X$ gauge coupling is much larger compared to the kinetic mixing parameter. However, in our work we choose $U(1)_X$ gauge coupling to be small enough so that such late conversion between two DM components does not happen.
For our choices of couplings and region of interest from XENON1T excess point of view, both DM-SM interactions as well $\chi_2 \chi_2 \rightarrow \chi_1 \chi_1$ freeze out at same epochs. Similarly, the interaction $\chi_2 e \rightarrow \chi_1 e$ also freezes out around the same epoch as other processes.

For a quantitative comparison, we estimate the cross sections of different processes relevant for both fermion and scalar DM masses below 100 MeV. For fermion DM, they are given by
\begin{eqnarray}
\sigma (\chi_2 \chi_2 \rightarrow \chi_1 \chi_1) & \ = \ & \frac{g'^4}{1536 \pi M^4_{Z'} s (s-4M^2_2)} f_1 (M_1, M_2, s, M_{Z'}, m_-) \nonumber \\
\sigma (\chi_2 e \rightarrow \chi_1 e) & \ = \ & \frac{g'^2 g^2 \epsilon^2}{8\pi \left ( (s-M^2_2-m^2_e)^2-4m^2_e M^2_2 \right)} f_2 (M_1, M_2, s, M_{Z'})\nonumber \\
\sigma (\chi_{2} \chi_{1} \rightarrow \nu \bar{\nu}) & \ = \ & \frac{g^2 g'^2 \epsilon^2 \left ( 2s + (M_1+M_2)^2 \right)}{48 \pi \left ( s - (M_1+M_2)^2 \right) (s-M^2_{Z'})^2} f_3(M_1,M_2,s) \nonumber \\
\sigma (\chi_{1,2} \chi_{1,2} \rightarrow \nu \bar{\nu}) & \ = \ &\frac{g^2 g'^2 \epsilon^2 m^2_-s}{96 \pi M^2_{1,2} (s-M^2_{Z'})^2} \sqrt{1-\frac{4 M^2_{1,2}}{s}} \nonumber\\
\end{eqnarray}

For scalar DM they are given by
\begin{eqnarray}
\sigma(\eta_2 \eta_2 \rightarrow \eta_1 \eta_1) &\ = \ & \frac{g'^4 s \sqrt{\frac{(s-4M^2_1)(s-4M^2_2)}{s^2}}}{16 \pi s (s-M^2_2) M^4_{Z'}}f_4(M_1,M_2,M_{Z'},s)\nonumber\\
\sigma(\eta_2 e \rightarrow \eta_1 e) & \ = \ & \frac{\epsilon^2 g'^2 g^2}{32 \pi \cos^2 \theta_w (s-M^2_2)^2}f_5(M_1,M_2,M_{Z'},s)\nonumber\\
\sigma(\eta_2 \eta_1 \rightarrow e^+ e^-) & \ = \ & \frac{\epsilon^2 g'^2 g^2 (M^2_1-s)(s-M^2_2)}{192 \pi \cos^2 \theta_w s^3 (s-M^2_{Z'})^2}f_6(M_1,M_2,M_{Z'},s)\nonumber\\
\sigma(\eta_2 \eta_1 \rightarrow \nu \overline{\nu}) & \ = \ & \frac{\epsilon^2 g'^2 g^2 s \sqrt{\frac{M^4_1+(s-M^2_2)-2M^2_2(s+M^2_2)}{s^2}}}{96 \pi \cos^2 \theta_w (s-M_{Z'})^2}
\end{eqnarray}

\begin{figure}[h!]
\centering
\includegraphics[scale=0.50]{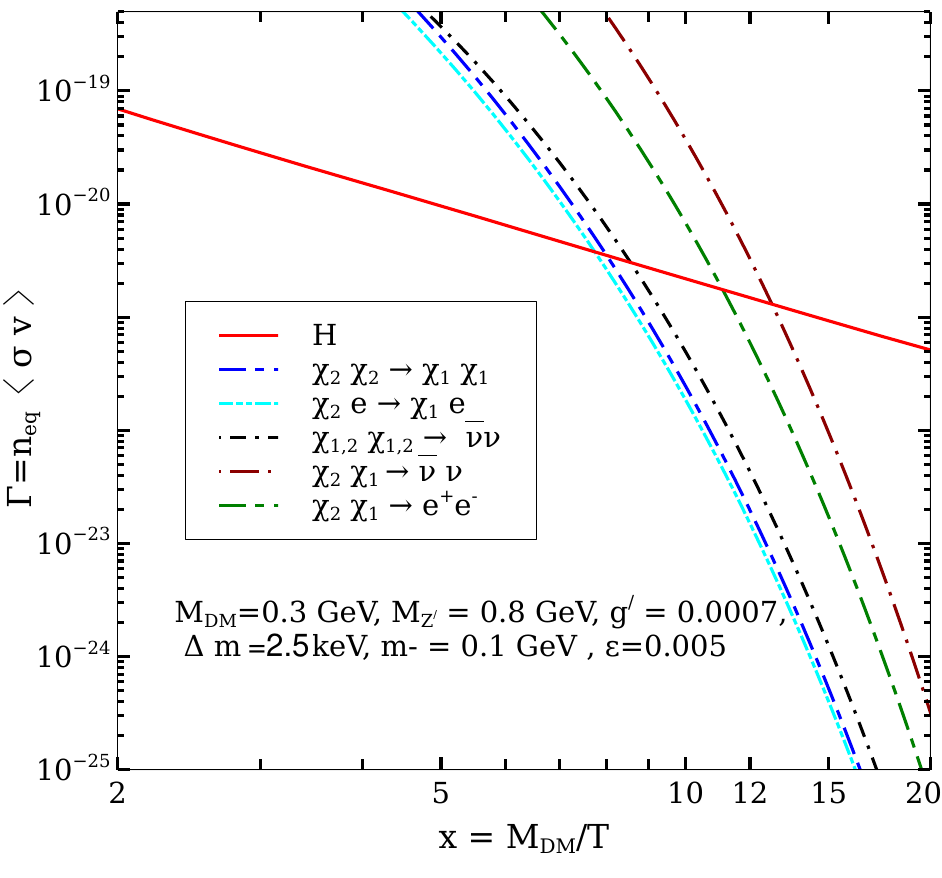}
\includegraphics[scale=0.50]{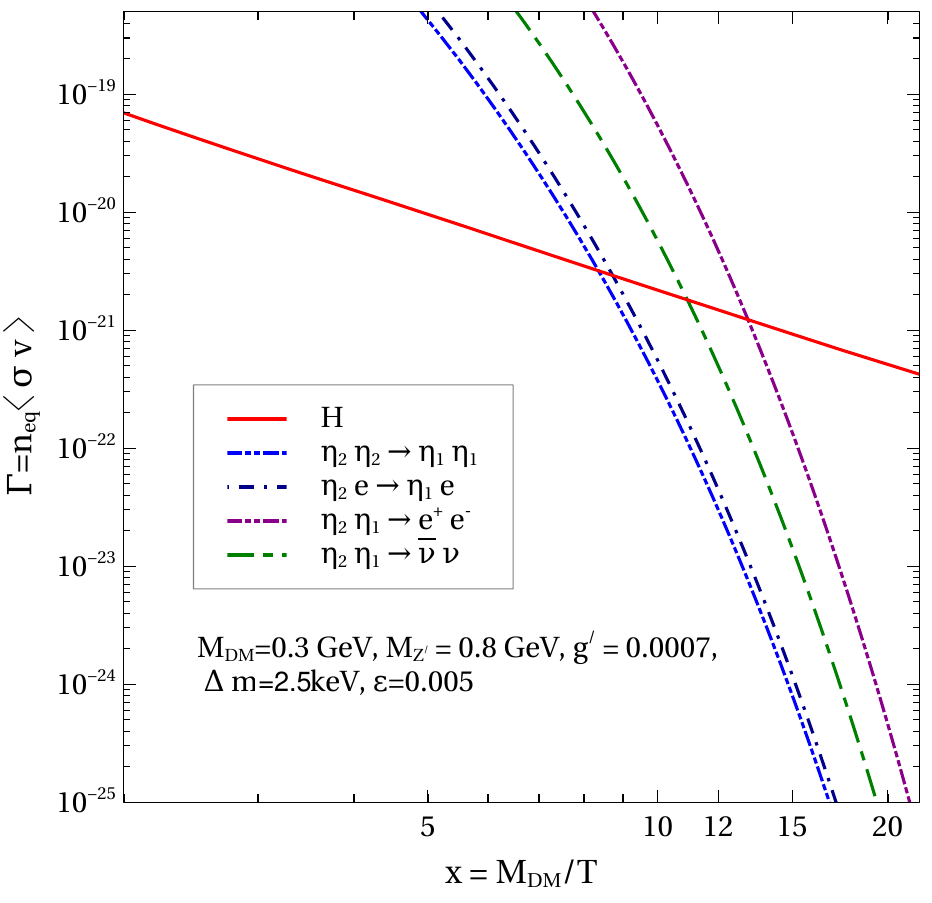}
\caption{Rates of different DM scattering processes in comparison to the Hubble expansion rate. Left (right) panel shows rates of different processes for fermion (scalar) DM.}
\label{Fig_decoupling}
\end{figure}

In the above expressions for cross sections $f_1, f_2, f_3, f_4, f_5, f_6$ are functions of model parameters, the details of which are skipped here for simplicity, but taken into account in the numerical calculations. It is important to note that the cross section of the first process (for both fermion and scalar DM) depends upon $U(1)_X$ gauge coupling $g'$ while the latter ones depend on $\epsilon$ as well. Clearly, for our chosen values of $g', \epsilon$ we have similar $g'^4$ and $g'^2 g^2 \epsilon^2$ where $g$ is the electroweak gauge coupling. For a comparison, we show the rates of these processes in comparison to Hubble expansion rate in figure \ref{Fig_decoupling}. We have used $g' = 0.0007, \epsilon=0.005, M_{\rm DM} = 0.3$ GeV, $ \Delta m = 2.5$ keV, $M_{Z'} = 0.8$ GeV. Clearly, the internal DM conversion processes decouple almost simultaneously with the DM annihilation and coannihilation processes, as expected. Therefore, the estimate of DM abundance based on the chemical decoupling is justified in our setup.

\subsection{Constraints from indirect detection}
\begin{figure}
\centering
\includegraphics[scale=0.6]{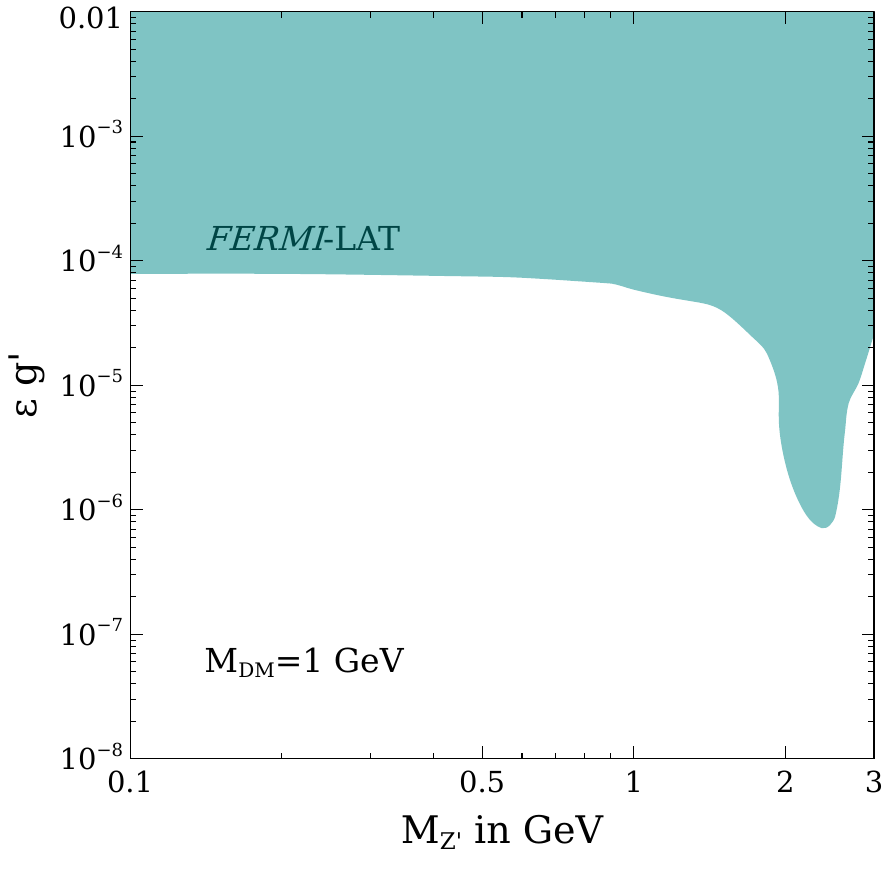}	
\caption{Constraints on the model parameters from indirect detection limits on DM annihilation rates into $\bar{u}u$. The grey shaded region is disfavoured 
from Fermi-LAT limits on DM annihilation rate to $\bar{u}u$.}
\label{inddetect}
\end{figure}

In addition to the relevant constraints on such sub-GeV inelastic DM, one can also have bounds on DM parameter space from indirect DM detection experiments which are searching for SM particles produced either through DM annihilations or via DM decay. We have already discussed DM decay in the context of heavier DM component's lifetime and showed that the requirement for heavier DM's lifetime to be larger than the age of the universe can be satisfied. However, constraints from indirect detection can be more restrictive as we discuss below. Here we consider both DM annihilation and decay into SM particles. Among such SM particles, photons being neutral and stable, can reach the indirect detection experiments without getting obstructed much in the intermediate regions. For DM in GeV regime or above, such photons produced from DM annihilations lie in the gamma ray regime and hence can be detected at space based telescopes like Fermi-LAT or ground based telescopes like MAGIC. Due to non-observations of any such gamma ray excess from dSphs at these experiments \cite{Ackermann:2015zua, Ahnen:2016qkx}, one can constrain the DM parameter space from GeV to tens of TeV mass regime. As these studies give constraints on DM masses from 1 GeV to tens of TeV only, we check for a benchmark DM mass of 1 GeV, the constraints imposed by these limits on our model parameter space. Note that the indirect detection limits on DM annihilation rates to specific final states are imposed assuming the final state to be $100\%$. In figure \ref{inddetect}, we show the constraints on parameter space of the model due to indirect detection limits on DM annihilation into 
up type quarks for DM mass of 1 GeV. Since in our model, DM annihilates to other lighter fermions as well, the bound shown in figure \ref{inddetect} should be considered as conservative upper bound only. As can be noticed by comparing this figure with our summary plot in figure \ref{fermionsummary}, indirect detection bound, even the most conservative one, does not rule out new region of parameter space of our interest. It should be noted that gamma ray constraints from above mentioned experiments are applicable for DM mass above 1 GeV only. A recent analysis \cite{Cirelli:2020bpc} have constrained such sub-GeV DM from INTEGRAL X-ray constraints. However, the bounds derived in this analysis are automatically satisfied by DM parameter space which survives other relevant bounds discussed in our work.

\begin{figure}
\centering
\includegraphics[scale=0.6]{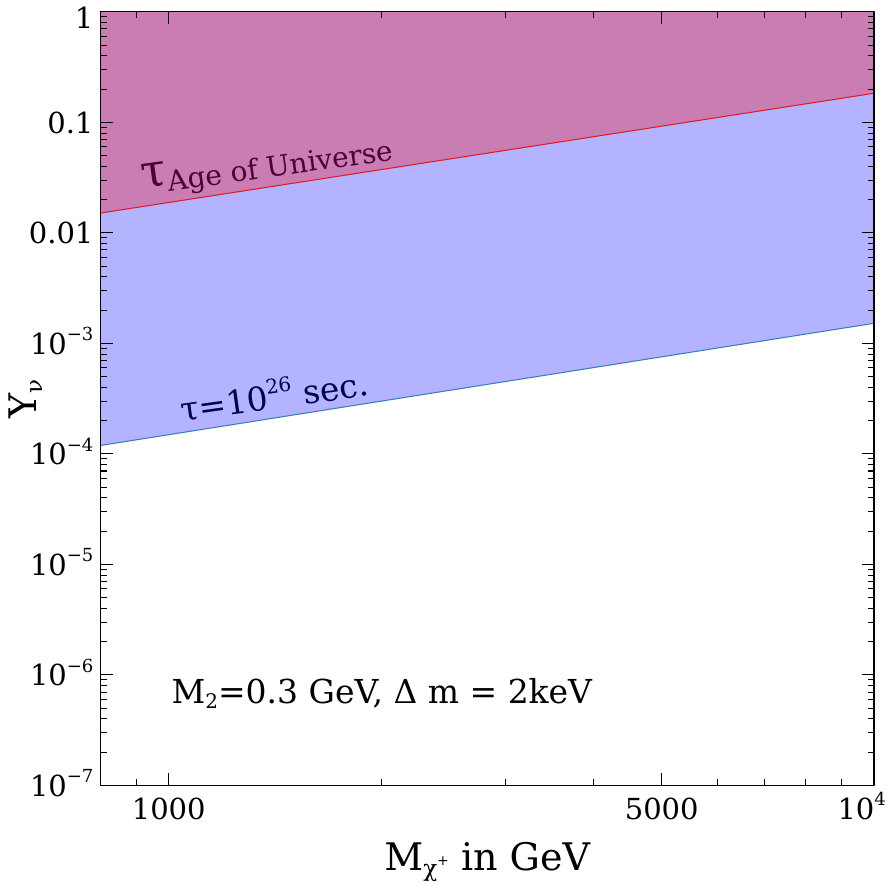}	
\caption{Constraints on the model parameters from lifetime limit on fermion DM decay mode $\chi_2 \rightarrow \chi_1 \gamma$ in radiative seesaw model.}
\label{inddetect1}
\end{figure}

Similar to DM annihilation into SM particles, DM decay can also be constrained from indirect detection experiments. In the models discussed here, the heavier DM decay into lighter DM and SM particles can offer such a possibility. For example, we can have a one-loop decay of heavier fermion DM as $\psi_2 \rightarrow \psi_1 \gamma \gamma $ which can lead to diffuse X-ray at energies below 2.5 keV. The corresponding decay width is however several order of magnitudes smaller compared to the tree level decay width discussed earlier, due to additional electromagnetic vertices and loop suppression. On the other hand, constraints on lifetime of DM decaying into photons is much stronger compared to the age of the universe. Typically, such lifetimes are constrained to be $\tau > 10^{26}$ s \cite{Essig:2013goa}. However, it is also worth noting that the existing bounds for diffuse X-ray photons at energies below 4 keV are rather weak, keeping our scenario safe \cite{Essig:2013goa}. While in most of the models discussed here, the dominant electromagnetic decay of heavier fermion DM leads to lighter DM and two photons in final states, in the radiative seesaw model, we can have the possibility of single photon final state also. To be more specific, it is possible to construct a one loop decay diagram of heavier fermion DM into lighter fermion DM and a photon using the same $ Y_{\nu} \overline{L} \tilde{\chi} \Psi_R $ vertex shown in the neutrino mass diagram in figure \ref{Fig0a}. Using the analysis of \cite{Pal:1981rm}, one can make a simple estimate of the corresponding decay width as 
\begin{equation}
\Gamma (\chi_2 \rightarrow \chi_1 \gamma) \approx \frac{e^2}{16\pi^5} (\Delta m)^3 Y^4_{\nu} \frac{M^2_2}{M^4_{\chi^+}} 
\end{equation}
where $M_{\chi^+}$ is the mass of charged component of the additional scalar doublet introduced in the radiative neutrino mass model. While this two body decay width goes as third power of mass splitting, compared to the three body decay width which depends upon fifth power of mass splitting as discussed before and hence can be dominant, we can choose the Yukawa coupling $Y_{\nu}, M_{\chi^+}$ appropriately to satisfy the lifetime bounds. Even though the experimental bounds on such DM decay into photons with energy 2.5 keV are rather weak, we can show that it is possible to have lifetime several order of magnitudes larger compared to the age of the universe \cite{Essig:2013goa}, for this two body decay involving a photon in final state. The corresponding parameter space consistent with age of universe bound as well as a conservative lower bound on lifetime $\tau > 10^{26}$ s is shown in figure \ref{inddetect1}. While the Yukawa coupling $Y_{\nu}$ also appears in neutrino mass generation, we can satisfy the limits from neutrino mass by suitable tuning of other free parameters present in the one-loop neutrino mass formula discussed earlier.
Similarly, the heavier scalar DM can also decay into the lighter DM and three photons at radiative level. However, due to additional loop and phase space suppressions in the corresponding decay width, lifetime of this decay $\eta_2 \rightarrow \eta_1 \gamma \gamma \gamma$ remains safe from diffuse X-ray bounds.

\subsection{Inelastic DM with large mass splitting}	
\begin{figure}[h!]
	\centering
	\includegraphics[scale=0.7]{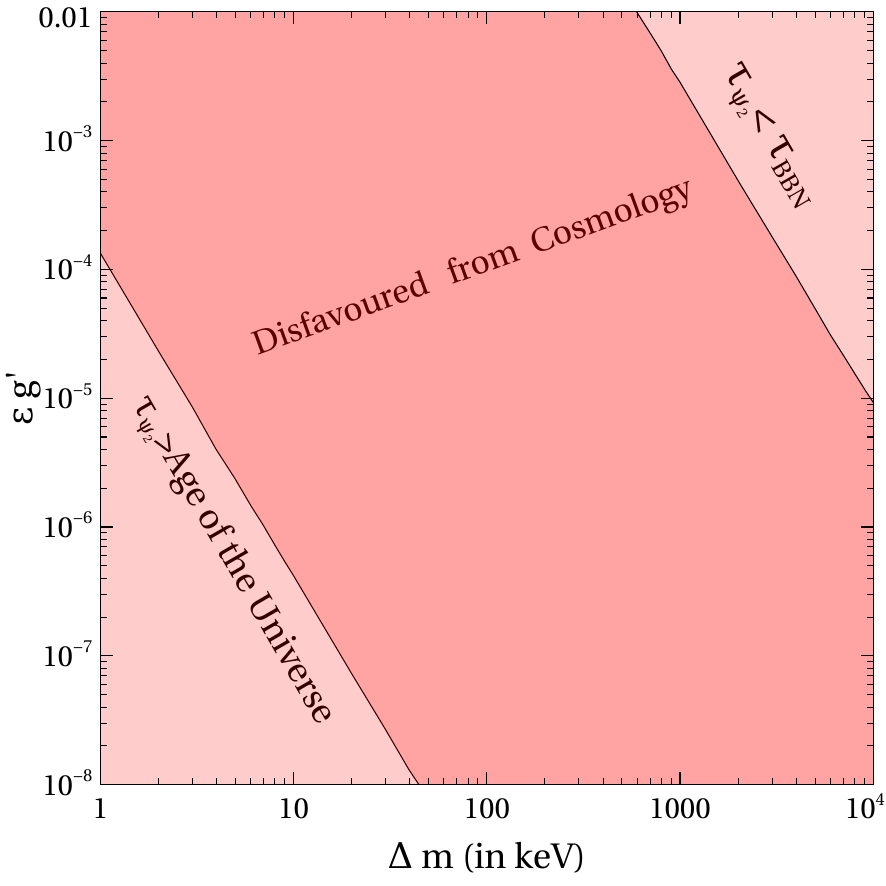}
	\caption{DM mass splitting versus $U(1)_X$ portal coupling showing the bounds on lifetime on heavier DM component.}
	\label{fig:life}
\end{figure}

\begin{figure}[h!]
	\centering
	\includegraphics[scale=0.6]{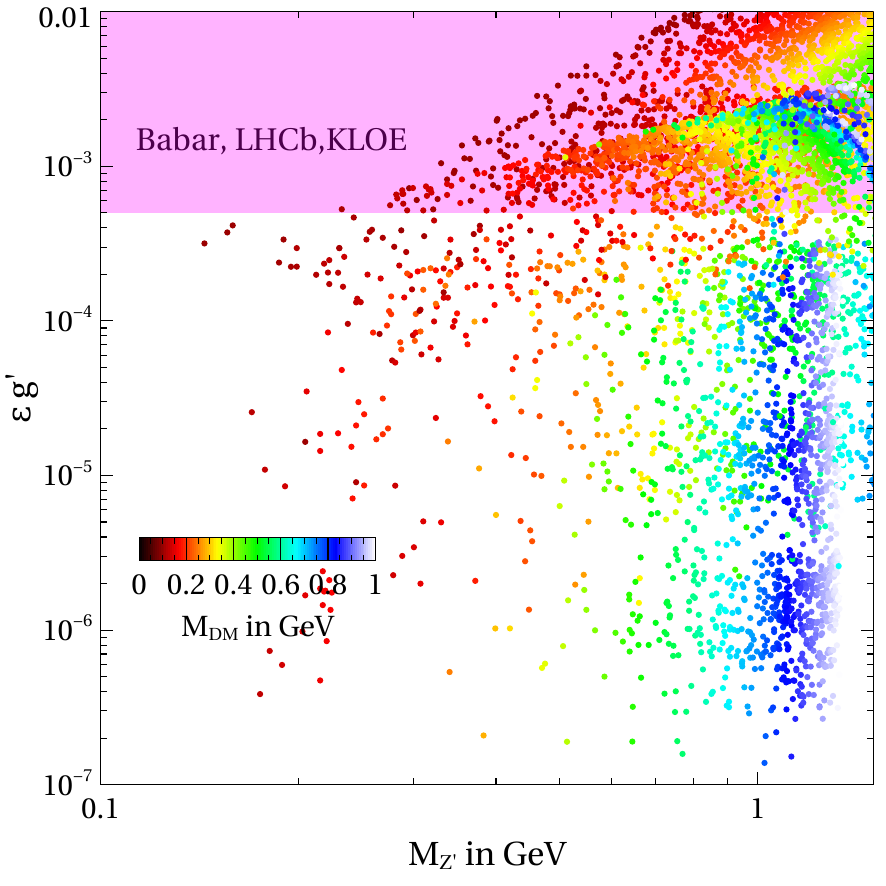}
	\caption{Summary plot for fermion DM with mass splitting of 1 MeV with next to lightest dark sector particle.}
	\label{fig:1MEV}
\end{figure}

It should be noted that we considered keV scale mass splitting between DM candidates in the above discussion. While such a scenario is preferred from XENON1T anomaly point of view, one can have correct DM phenomenology for other splitting also. For keV scale mass splittings, the decay width of heavier DM remains small enough to make it long lived compared to the age of the universe. However, if we increase the mass splitting, heavier DM can become unstable. Since heavier DM decays into lighter DM and other SM particles within kinematical limits, the lifetime will be constrained from non-observation of such decay products. We apply a conservative upper bound on lifetime of such unstable particle to be approximately equal to the epoch of big bang nucleosynthesis (BBN) such that late decay of such particles do not affect the successful predictions of BBN and other late events like recombination. The resulting parameter space is shown in figure \ref{fig:life}. Clearly, mass splitting from a few tens of keV to a few tens of MeV are disallowed from such lifetime bounds. Since the mass splitting also affects neutrino mass, the parameters involved in neutrino mass formula needs to be adjusted accordingly so that they are consistent with lifetime bounds as well as neutrino mass constraints. For large mass splitting say, 1 MeV, the DM sector comprises of the lighter component only as the heavier DM decays in the early epochs. While such a scenario can not explain the XENON1T excess, it is still a viable scenario from DM point of view. We show the allowed parameter space for sub-GeV scalar DM with such large mass splitting of 1 MeV in figure \ref{fig:1MEV}. While the overall pattern of scattered points remains similar, the constraints become much more relaxed. For example, there is no lifetime constraint in this case as heavier DM has already decayed prior to BBN epoch. Also, since mass splitting is large, DM can not up-scatter off a nucleon. Thus, direct detection bounds from experiments like CRESST do not apply here. We however, show the bounds from flavour physics experiments which rule out the upper portion of the parameter space.

Other cosmology bounds on such scenario may arise due to late decay of $Z'$ into SM leptons. For example, if $Z'$ decays after neutrino decoupling temperature $T^{\nu}_{\rm dec} \sim \mathcal{O}(\rm MeV)$, it will increase the effective relativistic degrees of freedom which is tightly constrained by Planck 2018 data as ${\rm N_{eff}= 2.99^{+0.34}_{-0.33}}$ \cite{Aghanim:2018eyx}. As pointed out by the authors of \cite{Ibe:2019gpv}, such constraints can be satisfied if $M_{Z'} \gtrsim \mathcal{O}(10 \; {\rm MeV})$, already satisfied by our scenario. Similar bound also exists for thermal DM masses in this regime which can annihilate into leptons till late epochs. Such constraints from the BBN as well as the CMB measurements can be satisfied if $M_{\rm DM} \gtrsim \mathcal{O}(1 \; {\rm MeV})$ \cite{Sabti:2019mhn} which is also satisfied in our models. 
	
\section{Conclusion}
\label{sec4}
Motivated by the recently reported excess of electron recoil events by the XENON1T experiment, we have studied the possibility of connecting low scale seesaw models for light neutrino masses with sub-GeV inelastic dark matter within the framework of Abelian gauge extension of the standard model. The vacuum expectation value of a scalar singlet field not only plays a crucial role in generating light neutrino masses but also splits the DM field into two quasi-degenerate components. In the limit of vanishing mass splitting between the two DM components, light neutrino mass also tends to zero thereby making the inelastic nature of DM a primary requirement from neutrino mass constraints. If the mass splitting is sufficiently small, say of the order of keV or below, both the DM components can be present in the universe as the lifetime of heavier DM can exceed the age of the universe for suitable choice of parameters. Interestingly, such keV scale mass splitting of sub-GeV DM can give rise to the electron recoil excess recently reported by the XENON1T collaboration.

From minimality point of view, we choose the Abelian gauge symmetry to be dark so that none of the SM particles are charged under it. DM particles can annihilate into SM particles through kinetic mixing of $U(1)_X$ and $U(1)_Y$. We constrain the parameter space of the model from the requirement of correct relic abundance as well as reproducing the XENON1T excess. We also impose bounds on direct detection cross section for such sub-GeV DM. We find that for sub-GeV DM with keV mass splitting the constraints coming from lifetime criteria on heavier DM component is the strongest while bounds from direct detection, flavour and LEP constraints are much weaker. The CMB bounds on DM annihilation to charged leptons, however, put a stricter bound on the parameter space, for DM masses beyond 1 GeV. We show that for both scalar and fermion inelastic DM, the desired phenomenology can be achieved. We also show that for our choices of model parameters, the late conversion within two DM candidates remain suppressed and hence our estimate of equal relic abundance of two DM candidates having keV mass splitting based on their chemical decoupling remains valid. We also check the constraints from indirect detection experiments on DM annihilation as well as DM decay into SM particles like charged fermions, photons and find that the parameter region of our interest remains safe from these bounds.

While keV mass splitting is crucial to fit XENON1T excess via down-scattering of heavier DM off electrons, heavier splittings of 1 MeV or so are also allowed. However, in such a case, only the lighter DM component is present in the universe while the heavier one decays before the epoch of BBN, keeping the relevant cosmological predictions undisturbed. The allowed DM specific parameter space in such a case becomes much more relaxed as lifetime bound of heavier DM as well as direct detection bounds are no longer applicable. Only the bounds from flavour physics experiments rule out some portion of the parameter space in such a scenario. Due to the presence of particle spectra around the TeV corner, the models proposed in this work can be tested in several experiments. On the cosmic frontier also, the model can have interesting implications like late decay of heavier DM component, enhanced annihilation rates into SM fermions which can be tested with CMB data. We leave a detailed study of such phenomenology for future studies.

\acknowledgements
DB acknowledges the support from Early Career Research Award from DST-SERB, Government of India (reference number: ECR/2017/001873). SM thanks Dibyendu Nanda and Shilpa Jangid for useful discussions. SM would also like to acknowledge Benjamin Roberts for useful discussions regarding the atomic ionisation factor.

\bibliographystyle{JHEP}
\bibliography{ref.bib}

\end{document}